\documentclass[a4paper,12pt]{article}
\pdfoutput=1
\usepackage{a4wide}
\usepackage[latin1]{inputenc}
\usepackage{dsfont}
\usepackage{amsfonts}
\usepackage{amsmath}
\usepackage{amssymb}
\usepackage{amsthm}
\usepackage{graphicx}
\usepackage[numbers, sort]{natbib}
\allowdisplaybreaks[1]
\usepackage{color}
\usepackage{ulem}
\usepackage{multirow}

\usepackage[small,bf]{caption}
\newcommand{\be}{\begin{equation}}
\newcommand{\ee}{\end{equation}}
\newcommand{\beq}{\begin{equation}}
\newcommand{\eeq}{\end{equation}}
\newcommand{\bea}{\begin{eqnarray}}
\newcommand{\eea}{\end{eqnarray}}

\newcommand{\braket}[1]{\langle #1 \rangle}

\usepackage{commath}
\usepackage{mathcomp}
\usepackage{colortbl}
\usepackage{bbold}
\usepackage{commath}

\newcommand{\Hubble}{ \mathcal{H} }

\newcommand{\emptyline}{ \vspace{15pt} }
\newcommand{\kahler}{K\"{a}hler }

\usepackage{slashed}

\usepackage[section, below, above]{placeins}

\definecolor{nochargecolor}{rgb}{1.0,1.0,1.0}
\newcommand{\nocharge}{ {\color{nochargecolor} 0} }

\newcommand{\da}{\Delta a}
\newcommand{\db}{\Delta b}

\begin{document}

\begin{titlepage}

\vspace*{-15mm}
\vspace*{0.7cm}

\begin{center}

{\Large {\bf Matter inflation with $A_4$ flavour symmetry breaking}}\\[8mm]

Stefan Antusch$^{\star\dagger}$\footnote{Email: \texttt{stefan.antusch@unibas.ch}} and
David Nolde$^{\star}$\footnote{Email: \texttt{david.nolde@unibas.ch}}

\end{center}

\vspace*{0.20cm}

\centerline{$^{\star}$ \it
Department of Physics, University of Basel,}
\centerline{\it
Klingelbergstr.\ 82, CH-4056 Basel, Switzerland}

\vspace*{0.4cm}

\centerline{$^{\dagger}$ \it
Max-Planck-Institut f\"ur Physik (Werner-Heisenberg-Institut),}
\centerline{\it
F\"ohringer Ring 6, D-80805 M\"unchen, Germany}

\vspace*{1.2cm}

\begin{abstract}
\noindent
We discuss model building in tribrid inflation, which is a framework for realising inflation in the matter sector of supersymmetric particle physics models. The inflaton is a D-flat combination of matter fields, and inflation ends by a phase transition in which some Higgs field obtains a vacuum expectation value. We first describe the general procedure for implementing tribrid inflation in realistic models of particle physics that can be applied to a wide variety of BSM particle physics models around the GUT scale. We then demonstrate how the procedure works for an explicit lepton flavour model based on an $A_4$ family symmetry. The model is both predictive and phenomenologically viable, and illustrates how tribrid inflation connects cosmological and particle physics parameters. In particular, it predicts a relation between the neutrino Yukawa coupling and the running of the spectral index $\alpha_s$. We also show how topological defects from the flavour symmetry breaking can be avoided automatically.
\end{abstract}
\end{titlepage}

\newpage

\section{Introduction}

The field of flavour physics is experiencing rapid experimental progress, e.g.\ increasingly precise measurements of the leptonic mixing angles and perspectives to measure currently unknown parameters, such as the leptonic Dirac CP phase and the neutrino mass ordering. This progress provides both a challenge for some of the existing models explaining the flavour structure of the Standard Model, and an opportunity to constrain and test such models with increasing precision. For example, the measurement of $\theta^{\text{PMNS}}_{13}$ by T2K~\cite{Abe:2011sj}, Double Chooz~\cite{Abe:2011fz}, RENO~\cite{Ahn:2012nd}, and in particular Daya Bay~\cite{An:2012eh}, has put tensions on popular models that predict tribimaximal mixing, and has encouraged the search for new models that are compatible with the experimental values \cite{reviews}.

A sufficiently complete particle physics model should also provide a consistent cosmological evolution. Many processes in the early universe happen at high energy scales, which makes these processes very sensitive to particle physics beyond the Standard Model. Inflation \cite{inf1, inf2, inf3, inf4} and baryogenesis are obvious examples. Inflation is driven by the potential of some scalar field, which can be identified with a new field of the particle theory, and baryogenesis is particularly sensitive to the flavour structure and CP violating interactions at high energies. Precision cosmology could therefore provide interesting hints to the high energy particle theory.

Several scenarios have been proposed for realising inflation with observable matter fields, e.g.\ Higgs inflation with non-minimal coupling to gravity \cite{smHiggsInflation1, smHiggsInflation2}, MSSM inflation near inflection points \cite{mssmInflation1,mssmInflation2}, and tribrid inflation \cite{tribrid1, gnsTribrid}. The inflaton in these models can be an observable gauge non-singlet field, like a Higgs field or some D-flat MSSM direction. In such models, the inflationary dynamics and the reheating phase depend on the couplings of these observable matter fields. Therefore, they have good prospects for connecting cosmological and particle physics observables, leading to more predictive, testable models.

In this paper, we discuss how tribrid inflation can be realised in flavour models. In the first section, we briefly review tribrid inflation and discuss the general procedure how it can be implemented in a given particle physics model. Afterwards, we provide an explicit example model with an $A_4$ family symmetry to demonstrate how this works in practice. The resulting model is predictive and phenomenologically viable. In particular, it connects cosmological and particle physics parameters, predicting a relationship between the running of the spectral index $\alpha_s$ and the neutrino Yukawa coupling $y_{\nu}$ or the right-handed neutrino mass $m_N$. In the last section we show that the model can automatically avoid the production of potentially dangerous topological defects at the end of inflation, because the waterfall field receives a small shift already during inflation. We then finish with a summary of our results.

\section{Tribrid inflation: A framework for connecting inflation with particle physics}

In this section, we discuss tribrid inflation \cite{tribrid1, gnsTribrid} and how it can be implemented in supersymmetric particle physics models. We start with a brief review and then describe the requirements and the strategy for realising tribrid inflation in a given model. We will later apply this general strategy to an explicit $A_4$ model in section~\ref{sec:A4model} as an illustrative example.

\subsection{Short review of tribrid inflation}
Tribrid inflation is a variant of supersymmetric hybrid inflation \cite{hybrid1, hybrid2, hybrid3}, which uses a superpotential of the schematic form
\begin{equation}
\label{eq:tribridW}
 W = \kappa S( H^\ell - M^2 ) + \lambda H^m \Phi^n,
\end{equation}
with $\ell \geq m \geq 2$. We use natural units with $M_{\text{Pl}} = (8\pi G)^{-1/2} = 1$ to keep the notation simple.

With this superpotential, tribrid inflation is defined as single-field inflation along the $\Phi$ direction while $H$, $S \simeq 0$. Inflation ends with a waterfall transition in $H$, which is triggered when $\lvert \Phi \rvert  <  \lvert \Phi_c \rvert$. The singlet field $S$ is merely an auxiliary field which is approximately zero during and after inflation.\footnote{$S$ usually gets a Hubble-size mass from the K\"{a}hler potential during inflation. For $m=2$, the auxiliary field is then stabilized at $S=0$. For $m>2$, $S \neq 0$ along the inflationary trajectory, but as long as it is heavy enough, we still have $S \ll \Phi$ and $\dot{S} \ll \dot{\Phi}$, which can be roughly approximated by $S \simeq 0$. \cite{pseudosmooth}} 

The main difference between tribrid inflation and conventional SUSY hybrid inflation, which uses the singlet $S$ as the inflaton, is that the tribrid inflaton $\Phi$ can be charged under symmetries. We can therefore use matter fields, e.g.\ D-flat MSSM directions, as inflaton directions. This connects inflation and particle physics more closely: inflation is most sensitive to the inflaton couplings, so if the inflaton is composed of observable matter fields, then the particle theory can more easily constrain inflationary predictions.

\emptyline

It turns out that depending on $\ell$ and $m$, the effective inflaton potential $V(\Phi)$ is dominated by different contributions, leading to qualitatively different regimes:
\begin{itemize}
 \item $\ell = m = 2$ (``loop-driven regime'') \cite{loopdriven1,loopdriven2,tribridHeisenberg}: In this case, $V(\Phi)$ is usually dominated by one-loop quantum corrections to the effective potential. The inflaton potential can receive dangerously large supergravity contributions from the K\"{a}hler potential, but these can be avoided either by the use of symmetries in the K\"{a}hler potential \cite{loopdriven1,tribridHeisenberg} or by a percent-level tuning of one K\"{a}hler potential parameter.
 \item $\ell \geq m = 2$ (``K\"{a}hler-driven regime'') \cite{kahlerdriven}: In this case, $V(\Phi)$ is generated mostly from Planck-suppressed operators in the K\"{a}hler potential.\footnote{For $\ell=2$, the loop potential is usually dominant (loop-driven regime), except for some specific K\"{a}hler potentials, where even $\ell=2$ can be K\"{a}hler-driven. For $\ell > 2$, loop corrections are strongly suppressed, and the K\"{a}hler-driven regime is the generic case.} For $\Phi \ll M_{\text{Pl}}$, where the effective field theory (EFT) is valid, the resulting inflaton potential usually has a hilltop-type shape with only three relevant parameters. K\"{a}hler-driven inflation can produce an observably large $\alpha_s > 0$, which -- if it should be measured -- could fix the most important model parameter, and also rule out the other regimes of tribrid inflation.
 \item $\ell \geq m > 2$ (``pseudosmooth regime'') \cite{pseudosmooth}: In this case, $V(\Phi)$ is generated from the superpotential coupling $\lambda H^m \Phi^n$, with $H \neq 0$ already during inflation. Inflation proceeds similar to smooth hybrid inflation, but ends with a waterfall. Note that because $H \neq 0$ already during inflation, no topological defects are formed during the waterfall transition. The predictions have so far been calculated for $\ell=m$ only, so it is not yet clear whether models with $\ell > m > 2$ can generate the observed CMB spectrum.
\end{itemize}
The generic predictions for tribrid inflation within the EFT regime are $r \lesssim 0.01$, $M \gtrsim O(10^{16} \, \text{GeV})$ and $\alpha_s \gtrsim 0$. The small tensor-to-scalar ratio is typical for small-field models and the mass scale $M \sim M_{\text{GUT}}$ is typical for hybrid inflation. Some of these are falsifiable predictions: an observable $\alpha_s > 0$ would exclude both the loop-driven and the pseudosmooth regimes, and an observable $\alpha_s < 0$ or $r > 0.01$ would exclude all three regimes of tribrid inflation.

In addition to these rough predictions, tribrid inflation provides relations between CMB observables and model parameters. These detailed predictions depend on the regime. For example, the K\"{a}hler-driven regime links the running of the spectral index $\alpha_s$ to the superpotential coupling $\lambda$ and the symmetry breaking scale $\braket{H}$. These relations can connect cosmology and particle physics in explicit models.

\subsection{Tribrid inflation model building}
\label{sec:tribridModelBuilding}

In this paper, we focus on tribrid inflation in the K\"{a}hler-driven regime:
\begin{align}
 \label{eq:kahlerTribridW}
 W = \kappa S( H^\ell - M^2 ) + \lambda H^2 \Phi^n,
\end{align}
with $\ell \geq 3$, $n \geq 2$.

For model building purposes, one should realize that $H^\ell$, $H^2$ and $\Phi^n$ can be D-flat combinations of different fields: $H^\ell \rightarrow H_1 H_2 \dots H_l$, $\Phi^n \rightarrow \Phi_1 \Phi_2 \dots \Phi_n$. We can, for example, use some D-flat MSSM direction as the inflaton by replacing $\Phi^3 \rightarrow L H_d E$ or $\Phi^2 \rightarrow L H_u$.

Note that $H^\ell$ and $H^m$ can be partially composed of different fields, as long as at least one of the fields in $H^\ell$ and $H^m$ is the same. For example, one could have
\begin{align}
 W_{\text{example}} = \kappa S( \underbrace{H_1 H_2 H_3}_{\hat{=} \, H^\ell} - M^2 ) + \lambda \underbrace{H_1 X}_{\hat{=} \, H^2} \Phi^n + ..., \label{eq:HlHmExample}
\end{align}
where $H_1$ is the component of the waterfall field that appears both in $H^\ell$ and $H^m$. As long as $H_2$ and $H_3$ get a large mass from the superpotential or the K\"{a}hler potential, the waterfall can only start along the $H_1$ direction, which is stabilized by $V_F \supset |\partial W / \partial X|^2 = |\lambda \Phi^{n} H_1|^2$ during inflation.\footnote{The full theory can contain extra terms, which we denoted with ``...'' in eq.~\eqref{eq:HlHmExample}, that fix the ratios of the $H_i$ to avoid massless directions in the $H_1 H_2 H_3$ hypersurface after inflation. An explicit example is given in section \ref{sec:A4model}.}

The field $X$ which appears only in $H^m$ and not in $H^\ell$ will not get a vacuum expectation value after inflation, and should therefore not be identified with a Higgs field, but with a conventional matter field like a right-handed neutrino.

\begin{figure}[tbhp]
  \centering$
\begin{array}{cc}
\includegraphics[width=0.48\textwidth]{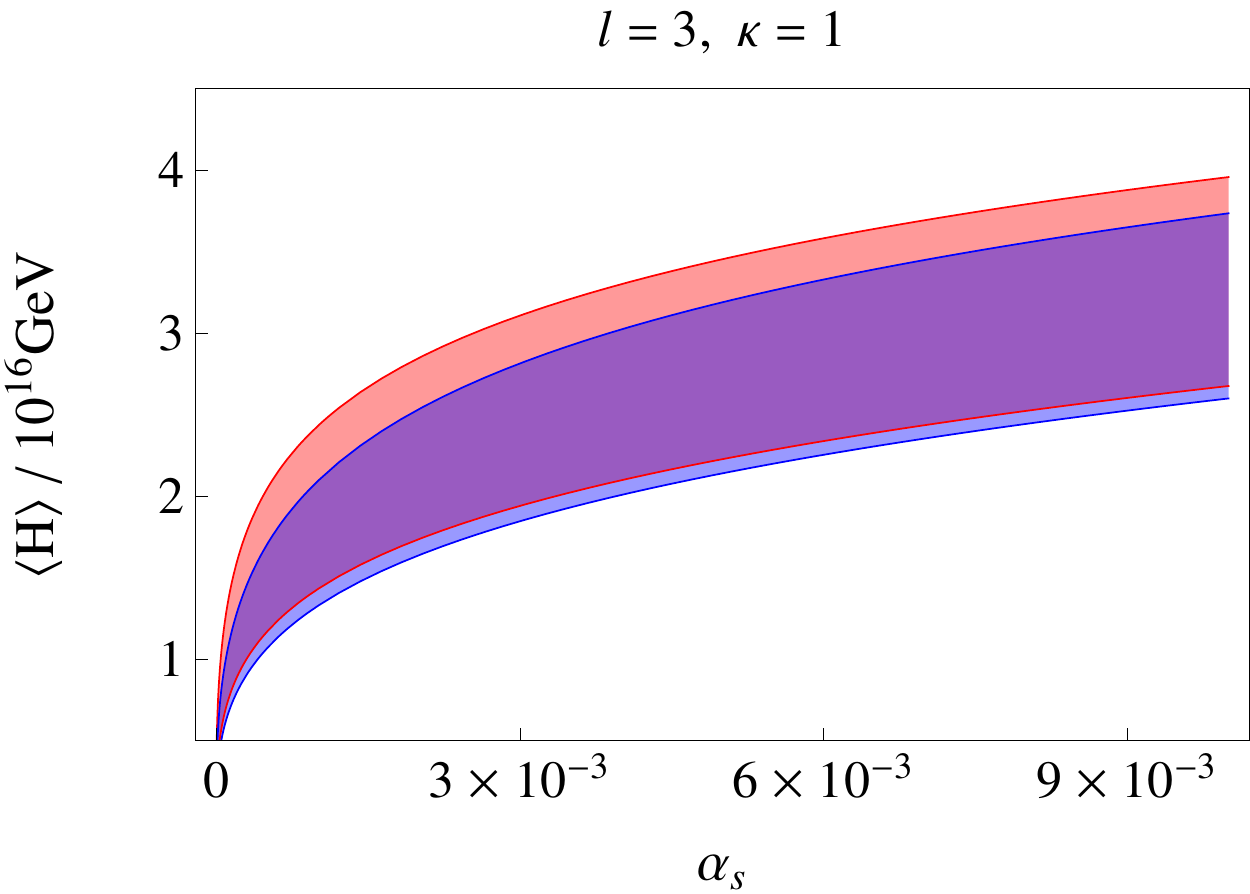} &
\includegraphics[width=0.48\textwidth]{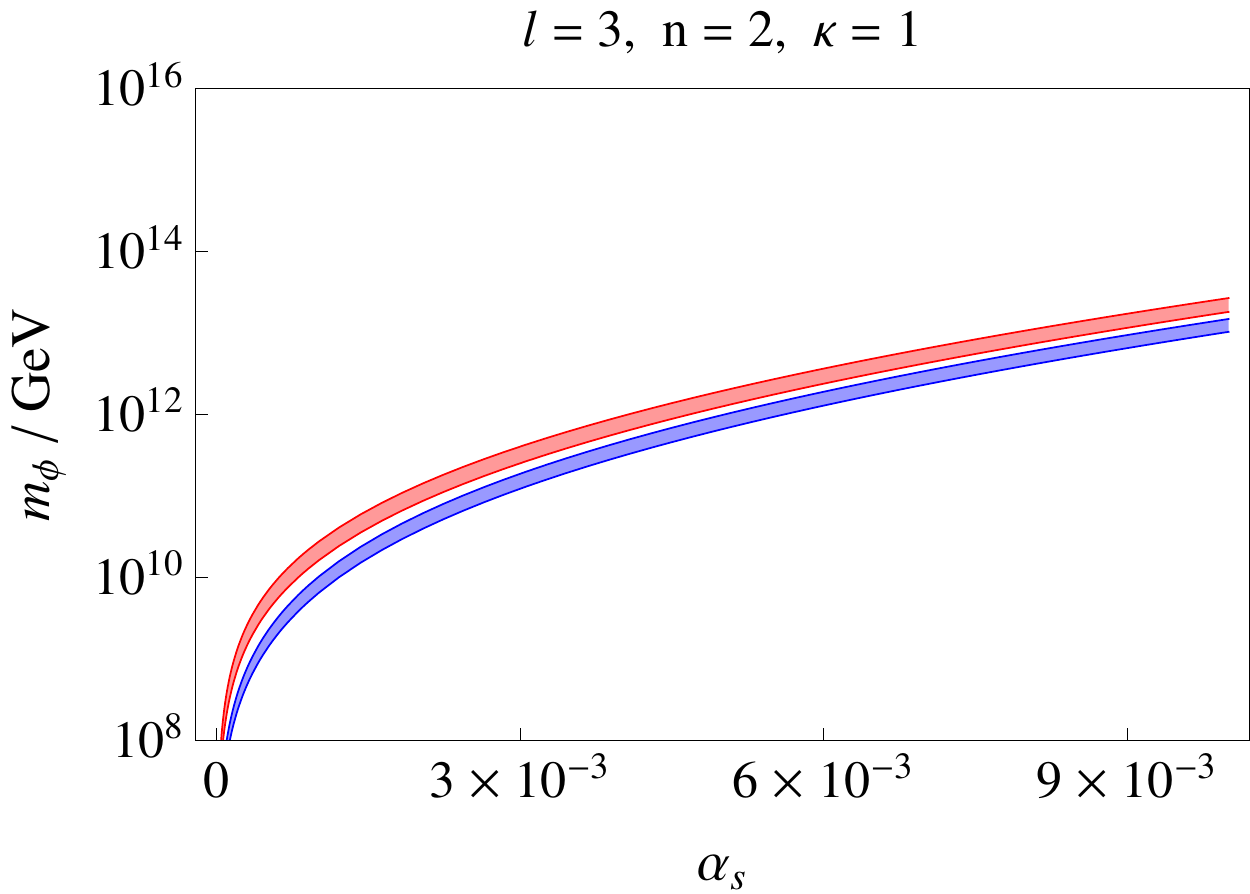} \\
\includegraphics[width=0.48\textwidth]{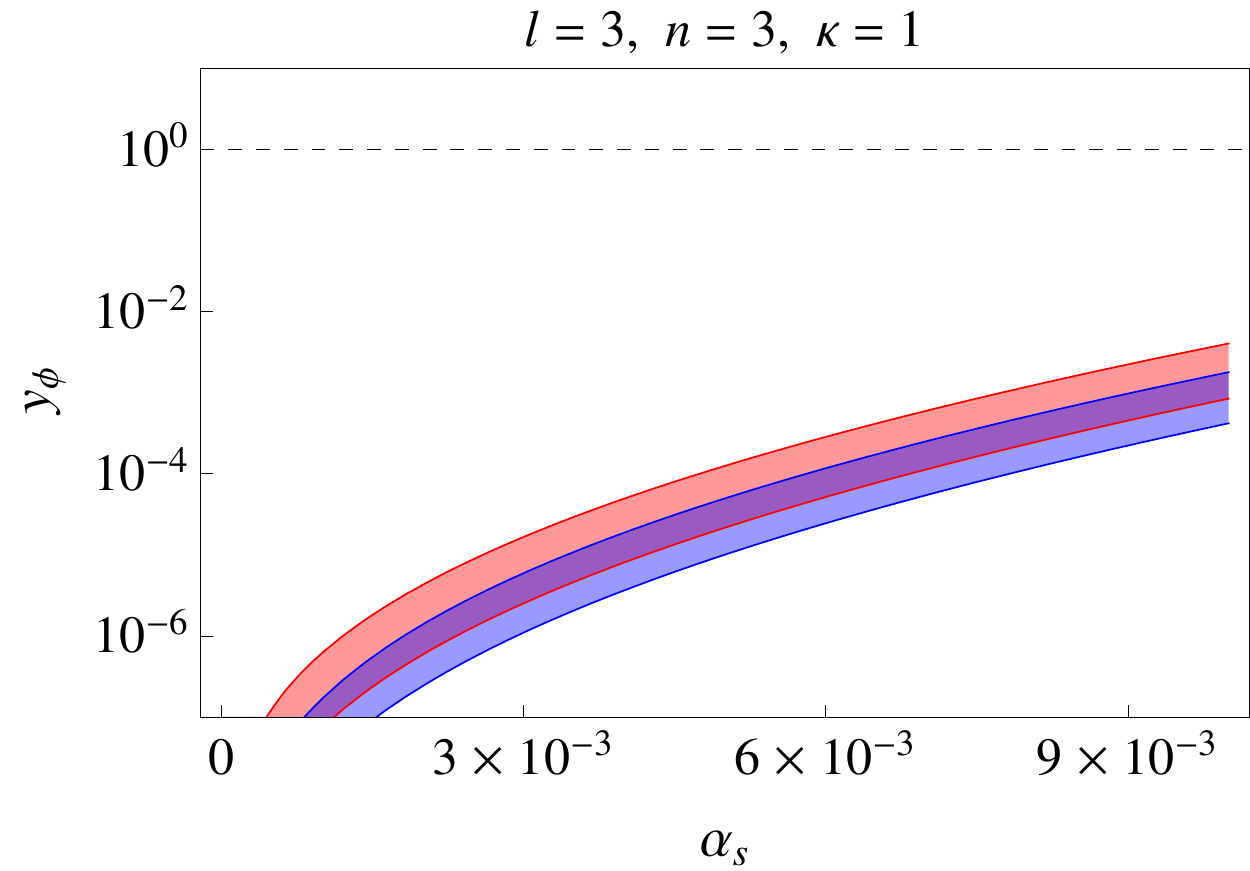} &
\includegraphics[width=0.48\textwidth]{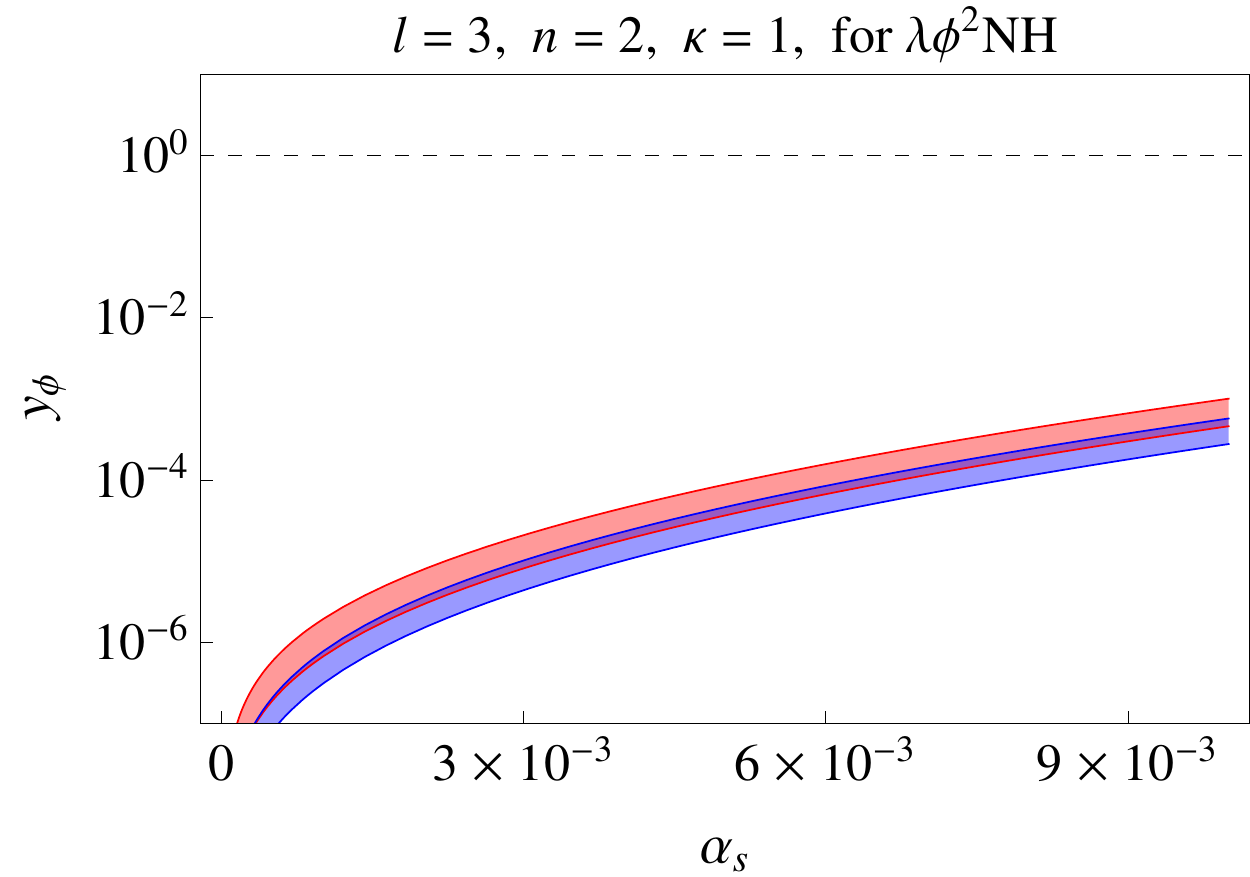}
\end{array}$
  \caption{Predicted values for the waterfall vacuum expectation value $\braket{H}$ and the predicted inflaton mass $m_\Phi$ or Yukawa coupling $y_\Phi$, for $n_s = 0.946$ (blue band) and $n_s = 0.975$ (red band), $N_0 = 55$, $\ell=3$, $\kappa = 1$ and $\kappa_{011} = 2$. For other values of $\kappa$ and $\kappa_{011}$, these predictions scale with $\braket{H} \propto \kappa^{-1/3}$ and $\lambda \propto (\kappa_{011} - 1)^{1/2}$.  The parameters $\ell,n,\lambda$ and $\kappa$ are defined in eq.~(\ref{eq:tribridW}), and $\kappa_{011}$ is a parameter of the K\"ahler potential as defined in ref.~\cite{kahlerdriven}. The width of the bands is due to free parameters of the K\"{a}hler potential related to the quartic inflaton coupling \cite{kahlerdriven}.}
  \label{fig:l3predict}
\end{figure}

After inflation, tribrid inflation predicts either a mass or a Yukawa coupling for the inflaton direction, depending on $n$ and on whether we use $H^2 \rightarrow H_1^2$ or $H^2 \rightarrow H_1 X$:

\begin{center}
\begin{tabular}{| c | c | c | c | }
  \hline
  Coupling term & $n$ & generated quantity & example application \\
  \hline
  \rule[-13pt]{0pt}{35pt}  $\lambda H_1^2 \Phi^2$ & $2$ & ${ \displaystyle  m_\Phi = 2\lambda \frac{\braket{H}^2}{M_{\text{Pl}}} }$ & neutrino Majorana mass term \\
  \rule[-13pt]{0pt}{35pt}  $\lambda H_1^2 \Phi^3$ & $3$ & ${ \displaystyle  y_\Phi = \lambda \left( \frac{\braket{H}}{M_{\text{Pl}}} \right)^2 }$ & quark or electron Yukawa coupling \\
  \rule[-13pt]{0pt}{35pt}  $\lambda H_1 X \Phi^2$ & $2$ & ${ \displaystyle  y_{\Phi X} = \lambda \frac{\braket{H}}{M_{\text{Pl}}} }$ & neutrino Yukawa coupling \\
  \hline
\end{tabular}
\end{center}
\emptyline

The examples mentioned in the table are
\begin{itemize}
 \item generating a neutrino Majorana mass term from $\lambda H_1^2 N^2$, where the right-handed sneutrino $N$ is the inflaton direction
 \item generating a quark or charged lepton Yukawa coupling from e.g.\ $\lambda H_1^2 L H_d E$, where $L H_d E$ is the inflaton direction
 \item generating a neutrino Yukawa coupling from e.g.\ $\lambda H_1 L H_u N$, where $L H_u$ is the inflaton direction and $N$ is zero during and after inflation.
\end{itemize}

The factor 2 in the table for the generated mass appears if a Majorana mass is generated. If one uses a composite inflaton direction $\Phi^2 \rightarrow \Phi_1 \Phi_2$, one instead generates a Dirac mass term, and the factor 2 is absent.

\subsubsection*{Linking model parameters to the CMB}
 
Slow-roll inflation links the model parameters to observables in the cosmic microwave background (CMB), so the parameters can be constrained from measurements of the spectral index $n_s$ and the running of the spectral index $\alpha_s$. The resulting constraints on $m_\Phi$ and $y_\Phi$ after inflation are shown in fig.~\ref{fig:l3predict}.\footnote{In \cite{kahlerdriven}, $\alpha_s$ is given as a prediction depending on the inflaton's \kahler potential coupling $a$ and the spectral index $n_s$. For this paper, we inverted $\alpha_s(a,n_s)$ and fixed the coupling $a$ by the observables $\alpha_s$ and $n_s$, which makes the plots more convenient for comparing tribrid models with observations.}

Inflation also constrains the symmetry breaking scale of the waterfall transition by constraining the value of $\kappa M^2$. This translates into bounds on $\braket{H}$, which are also shown in fig.~\ref{fig:l3predict}.

Note that $\kappa$ is a free parameter of the model. The plots are shown for $\kappa = 1$ and $\ell=3$. For smaller $\kappa$ and larger $\ell$, the predicted masses and Yukawa couplings are larger due to $\braket{H} = ( \sqrt{V_0} / \kappa )^{1/\ell}$, with $V_0$ fixed by inflation.

Similar plots for $\ell\neq 3$ can be found in appendix \ref{appendix:kahlerplots}.

\section{Tribrid inflation in an $A_4$ lepton flavour model}
\label{sec:A4model}
We now want to illustrate how tribrid inflation can be implemented in an explicit particle physics flavour model. We present an example model based on a discrete $A_4$ family symmetry\footnote{For early models with $A_4$ family symmetry, see e.g.\ \cite{Altarelli:2010gt}.} to generate the leptonic Yukawa couplings and neutrino masses. We will assume that in addition to the family symmetry, there is also a CP symmetry which is broken spontaneously only by the vacuum expectation values (vevs) of the flavons. One can easily realize tribrid inflation in this model using either the D-flat $LH_u$ direction or a sneutrino direction. The model then predicts a relationship between the running of the spectral index $\alpha_s$ and the neutrino Yukawa coupling or the heavy neutrino mass.

Interestingly, the production of topological defects during the waterfall transition may be avoided, because the waterfall field can be slightly shifted already during inflation. This is related to our choice of the flavon alignment potential, as we will discuss later in section \ref{sec:topologicalMechanism}.

Our model is an alternative to flavon inflation \cite{flavonInflation}, where one of the $A_4$ symmetry breaking flavons is the inflaton. Both models use similar types of superpotentials and relate the flavour symmetry breaking scale to the amplitude of the CMB fluctuations.\footnote{It has also been considered in \cite{flavonInflation} that one of the so-called driving fields from the flavon potential acts as the inflaton.} However, in our tribrid model the inflaton is a combination of matter fields. Since the CMB spectrum is related to the inflaton couplings, this can lead to particularly close relations between observables from inflation and particle physics. One advantage of using flavons as the inflaton \cite{flavonInflation} is that topological defects from the $A_4$ symmetry breaking are automatically inflated away, because the symmetry breaking occurs at the beginning and not at the end of inflation. Our model instead requires a mechanism to avoid or remove topological defects from the waterfall transition, like the one discussed in section \ref{sec:shiftedTribridA4}.

\subsection{Model outline}
The model is based on the superpotential
\begin{align}
\label{eq:eqW}
 W &= \lambda_i \, \Theta_i \cdot L H_u N_i  +  \gamma_{ij} (N_i \Theta_i) (N_j \Theta_j) \Theta_S  \notag \\
 &+ \lambda_4 \Theta_4 \cdot L H_d E_1  +  \lambda_5 \Theta_5 \cdot L H_d E_2  +  \lambda_6 \Theta_3 \cdot L H_d E_3  +  W_{\text{fl}}  +  W_{\text{misc}},
\end{align}
where summation over $i$, $j = 1$, $2$ is implied. The lepton doublets $L$ and the flavons $\Theta_n$ are $A_4$ triplets, while the flavon $\Theta_S$ is an $A_4$ singlet.

In this expression, the first terms are the lepton Yukawa couplings and the right-handed neutrino masses. $W_{\text{fl}}$ is the flavon alignment superpotential that enforces the correct breaking of the $A_4$ flavour symmetry via vacuum expectation values $\braket{\Theta_n}$. $W_{\text{misc}}$ contains the quark Yukawa couplings and possibly additional interactions which are not relevant for our purpose. How this model relates to the simple tribrid superpotential from eq.~\eqref{eq:kahlerTribridW} will be explained in section \ref{sec:inflatonDirections}.

The flavon alignment superpotential of our model has the following form: 
\begin{align}
 W_{\text{fl}} \, = \, &S_3 \left(  \Theta_3^{n_3} - \mu_3^2  \right)  +  S_4 \left(  \Theta_4^{n_4} - \mu_4^2  \right)  +  S_6 \left(  \Theta_6^{n_6} - \mu_6^2  \right)  +  S_7 \left(  \Theta_S^{n_7} - \mu_7^2  \right) \notag \\  
 &+  P_{34} \Theta_3 \Theta_4  +  P_{36} \Theta_3 \Theta_6  +  P_{46} \Theta_4 \Theta_6 
 +  A_3 \Theta_3 \star \Theta_3  +  A_4 \Theta_4 \star \Theta_4    \notag \\
 &+ S_5 \left(  [ \Theta_5 ]^6  -  \mu_5^2  \right)  +   D_5 \left\{  ( \Theta_5^2 )^2 + k_5 ( \Theta_5 \star \Theta_5 )^2  \right\}  + P_{35} \Theta_3 \Theta_5 \notag \\
 &+ S_1 \left(  \Theta_1^{n_1} - \mu_1^2  \right)  +  P_{16} \Theta_1 \Theta_6  \notag \\
 &+ \kappa S_2 \left(  \Theta_2 ^3 - \mu_2^2  \right)  +  \alpha_1 D_2'( \Theta_2^2 )_{1''}  +  \alpha_2 D_2''( \Theta_2^2 )_{1'} + P_{12} \Theta_1 \Theta_2. \label{eq:flavonW}
\end{align}
Terms $[\Theta_i]^n$ stand for all possible $A_4$ contractions of the field.\footnote{For products of $A_4$ representations, we use the Ma-Rajasekaran (``$SO(3)$-like'') basis, as introduced in the first reference of \cite{Altarelli:2010gt}.} $\Theta_2^3$ is the symmetric $A_4$ product of three triplets.
The symmetries and charge assignments for the entire superpotential are discussed in appendix \ref{appendix:symmetries}.
Prefactors which are not needed in the later calculations are dropped for brevity. $\kappa$, $\mu_2$, $\alpha_1$ and $\alpha_2$ are real in a suitable basis due to the assumed CP symmetry (before family symmetry breaking). $n_1$, $n_3$, $n_4$ and $n_6$ can take even integer values and $n_7$ any integer value greater than two.

The flavon alignment potential fixes the vacuum expectation values in the true vacuum to be
\begin{align}
 \braket{ \Theta_1 } &= c_1 \begin{pmatrix} 0 \\ 1 \\ \pm 1 \end{pmatrix}, &
 \braket{ \Theta_2 } &= c_2 \begin{pmatrix} \pm 1 \\ \pm 1 \\ 1 \end{pmatrix}, &
 \braket{ \Theta_3 } &= c_3 \begin{pmatrix} 0 \\ 0 \\ 1 \end{pmatrix}, &  \notag \\
 \braket{ \Theta_4 } &= c_4 \begin{pmatrix} 0 \\ 1 \\ 0 \end{pmatrix}, &
 \braket{ \Theta_5 } &=  \begin{pmatrix} c_5 \sin (\vartheta_5)\\ \mathrm{i} \,c_5 \cos (\vartheta_5)\\ 0 \end{pmatrix}, &
 \braket{ \Theta_6 } &= c_6 \begin{pmatrix} 1 \\ 0 \\ 0 \end{pmatrix}, &
 \braket{ \Theta_S } &= c_7,
\end{align}
where the $\pm$ can be chosen independently from each other with the constraint that $\braket{ \Theta_1 } \perp \braket{ \Theta_2 }$, and $c_i \in \mathbb{R}$, with global phases of the $ \braket{ \Theta_i}$ absorbed by redefinitions of the fields. 
For $k_5 > 3$ in eq.~\eqref{eq:flavonW}, the terms for $\Theta_5$ lead to a phase difference of $\pi/2$ between the two components of the flavon vev. The relative size of the components, parameterized by $\vartheta_5$, can be adjusted by the value of $k_5$ (cf.\ \cite{Antusch:2013kna}). 

For $\braket{ \Theta_2 } = c_2(1,1,1)^T$, this vacuum alignment will generate leptonic Yukawa matrices of the form
\begin{align}
 y_{\nu} = \begin{pmatrix}
            0 & \varepsilon_2 \\
            \varepsilon_1 & \varepsilon_2  \\
            -\varepsilon_1 & \varepsilon_2
            \end{pmatrix}, \quad
 y_{e} = \begin{pmatrix}
            0 & \varepsilon_5 \sin (\vartheta_5) & 0 \\
            \varepsilon_4 & \mathrm{i}\,\varepsilon_5 \cos (\vartheta_5) & 0 \\
            0 & 0 & \varepsilon_6
           \end{pmatrix},
\end{align}
and a diagonal right-handed neutrino mass matrix\footnote{The non-diagonal contribution from $\gamma_{12} \braket{ \Theta_S \Theta_1 \Theta_2 } N_1 N_2$ vanishes because $\braket{\Theta_1} \perp \braket{\Theta_2}$.}
\begin{align}
 M_R = \begin{pmatrix}
            M_1 & 0 \\
            0 & M_2
           \end{pmatrix}, \quad M_i = 2 \gamma_{ii} \braket{\Theta_i^2} \braket{\Theta_S}.
\end{align}
$y_\nu$ and $M_R$ generate the light neutrino masses via the seesaw mechanism and realize so-called ``constrained sequential dominance''~\cite{King:2002nf} where the mixing in the neutrino sector is tri-bimaximal \cite{Harrison:2002er}. 
The total leptonic mixing receives a contribution from the 1-2 mixing in $y_e$. This induces a non-zero $\theta_{13}^{\text{PMNS}}$ 
\begin{align}
 \theta_{13}^{\text{PMNS}} \simeq \left|\frac{\vartheta_5}{\sqrt{2}}\right| .
\end{align}
Furthermore, the leptonic Dirac CP phase $\delta^{\text{PMNS}}$ is maximal,
 \begin{align}
 \delta^{\text{PMNS}} \simeq \pm 90^\circ ,
 \end{align}
due to the phase difference between the components of $ \braket{ \Theta_5 } $. Plugging this into the lepton mixing sum rule \cite{sumrule}, 
$
 \theta_{12}^{\text{PMNS}} - \theta_{13}^{\text{PMNS}} \cos( \delta^{\text{PMNS}} ) \simeq 35.3^\circ 
$, 
which is applicable here since the 1-3 mixings in the mass matrix of the light neutrinos and in $y_e$ both vanish, we can further see that $ \theta_{12}^{\text{PMNS}}$ is not affected by charged lepton 1-2 mixing, and therefore is predicted to be close to the tri-bimaximal value. Also $\theta_{23}^{\text{PMNS}}$ does not receive a contribution from charged lepton mixing, and we obtain
\begin{align}
 \theta_{12}^{\text{PMNS}} \simeq 35.3^\circ , \quad \theta_{23}^{\text{PMNS}} \simeq 45^\circ\:.
\end{align}
Thus, $\theta_{12}$, $\theta_{23}$ and $\delta^{\text{PMNS}}$ are predicted by the model, while $\theta_{13}$ can be fitted by choosing a suitable $\vartheta_5$, which in turn depends on the parameter $k_5$ in the flavon alignment potential.\footnote{When a model of this type is embedded into a GUT, then the parameter $\vartheta_5$ is linked to the quark sector, which can lead to a prediction also for $\theta_{13}^{\text{PMNS}}$.}

\subsection{Inflaton directions for tribrid inflation}
\label{sec:inflatonDirections}

The first step for implementing tribrid inflation in any model is to find suitable inflaton and waterfall directions in the superpotential. In our $A_4$ flavour model, possible waterfall directions are given by the flavon directions $\Theta_i$ and possible inflaton directions are given by D-flat combinations of matter fields.

As we have mentioned in section \ref{sec:tribridModelBuilding}, K\"{a}hler-driven tribrid inflation requires a superpotential term of the form $\lambda H^2 \Phi^n$ or $\lambda H X \Phi^n$, where $H$ is the waterfall field, $\Phi^n$ is a D-flat inflaton direction and $X$ is some matter field with $X \simeq 0$ during and after inflation. In the superpotential \eqref{eq:eqW}, the candidate terms are
\begin{enumerate}
 \item $\lambda_j \, \Theta_j \cdot L H_u N_j$, with $H \rightarrow \Theta_j$, $\Phi^2 \rightarrow L H_u$ and $X \rightarrow N_j$,
 \item $\gamma_{jj} \braket{\Theta_S} N_j^2 \Theta_j^{2}$, with $H^2 \rightarrow \Theta_j^2$, $\Phi^2 \rightarrow N_j^2$,
\end{enumerate}
for any $j \in \{1,2\}$, without implied summation over $j$.

In the first case, $N_j=0$ during inflation, and tribrid inflation predicts the Yukawa coupling $y_{\nu_j} = \lambda_j \braket{ \Theta_j }$ as a function of $\alpha_s$ (fig.~\ref{fig:l3predict}). In the second case, the right-handed sneutrino $N_j$ is the inflaton and tribrid inflation predicts its mass $m_{N_j} = 2\gamma_{jj} \braket{\Theta_S} \frac{ \braket{\Theta_j^2} }{ M_{\text{Pl}}^2 }$ as a function of $\alpha_s$ (fig.~\ref{fig:l3predict}).

We assume that $\Theta_S$ and all $\Theta_i$ ($i=1,...,6$) except $\Theta_j$ have already settled at their minimum 60 e-folds before the end of inflation,\footnote{Otherwise, we would have multi-field inflation with several waterfalls. Only after the last waterfall will the vacuum energy be zero, so we expect that inflation continues until the last flavon field moves towards its minimum.} so that only the dynamics of the inflaton $\Phi^n$ and the chosen waterfall field $\Theta_j$ are important.

\subsubsection{Sneutrino inflation}

Using $N_j^2 =: \Phi^2$ as the inflaton direction is straightforward, as $N_j$ does not have any $A_4$ structure, and the operator $\braket{\Theta_S} \Theta_j^2 N_j^2$ stabilizes all components of the $A_4$ triplet flavon $\Theta_j$ during inflation. The relevant terms during inflation exactly reduce to the simple structure of eq.~\eqref{eq:kahlerTribridW}, and we expect that the usual results for K\"{a}hler-driven tribrid inflation apply.\footnote{The off-diagonal neutrino mass term $\gamma_{12} \Theta_S \Theta_1 \Theta_2 N_1 N_2$ can be ignored during inflation because $\Theta_1 \perp \Theta_2$ is enforced by the flavon alignment potential \eqref{eq:flavonW}.}

\subsubsection{Lepton-Higgs inflation}
\label{sec:LHuDirections}

Using $LH_u$ as the inflaton direction, we must account for the possibility that the $\braket{ \Theta_i }$, $i \neq j$, can induce a large F-term potential for the inflaton:
\begin{align}
 \left|  \frac{\partial W}{\partial N_i}  \right|^2  \, &= \,  \lvert  \lambda_i \Theta_i \cdot L H_u  +  2\gamma_{ii} \braket{\Theta_S} \Theta_i^2 N_i  +  \gamma_{ij} \braket{\Theta_S} \Theta_i \Theta_j N_j \rvert^2.
\end{align}
During inflation, these terms can generate a large inflaton potential except if $L \perp \braket{ \Theta_i }$.\footnote{The F-terms could also be minimised by $N_i = \lambda_i \Theta_i \cdot L H_u / ( 2 \gamma_{ii} \braket{\Theta_S} \Theta_i^2 )$. This ambiguity disappears if we include Hubble-scale mass terms for the $N_i$, which can be naturally generated from the K\"{a}hler potential.} This means that the inflaton component with $L \parallel \braket{\Theta_i}$ will be quickly driven to zero. If inflation lasts more than the bare minimum of 60 e-folds, we can therefore assume that $L \perp \braket{ \Theta_i }$ throughout the last 60 e-folds of inflation.

For $j=1$ or $j=3$, the model then again reduces to eq.~\eqref{eq:kahlerTribridW}, with $L H_u =: \hat{e}_\phi \Phi^2$, where $\hat{e}_\phi \perp \braket{\Theta_{i \neq j}}$ is the inflaton direction in $A_4$ space, normalized to $| \hat{e}_\phi |^2 = 1$. The case $j=2$ is more complicated and will be discussed below.

\subsubsection{Unique inflaton direction for preventing topological defects}
\label{sec:inflatonDirectionsTopological}

Many models of hybrid inflation can produce dangerous topological defects at the end of inflation, and the above model is no exception. In our case, the waterfall transition for the $\Theta_j$ usually generates $\mathbf{Z}_n$ domain walls. The model therefore needs some mechanism to either prevent their production or to remove the domain walls after they have formed.

In tribrid inflation, the problem is alleviated by the fact that the inflaton can be charged under the symmetries of the waterfall field, and therefore those symmetries are usually broken already during inflation. If the symmetry breaking is transmitted to the waterfall field, e.g.\ via additional inflaton-waterfall field couplings, the formation of topological defects at the end of inflation may be avoided. \cite{gnsTribrid}

However, in this paper's $A_4$ model there is one unique inflaton direction for which the production of domain walls is avoided automatically: the $LH_u$ direction with $\Theta_2$ as the waterfall field ($j=2$). We want to discuss this case in more detail in section~\ref{sec:shiftedTribridA4}.

\subsection{Summary}
In this section, we have presented an explicit flavour model based on an $A_4$ symmetry, which makes testable predictions for $\theta_{12}^{\text{PMNS}}$, $\theta_{23}^{\text{PMNS}}$ and for the Dirac CP phase $\delta^{\text{PMNS}}$. We have shown that tribrid inflation can be easily realised in these models: we found six possible inflaton trajectories which predict a relation between the running of the spectral index $\alpha_s$ and either one of the neutrino Yukawa couplings $y_\nu$ or one of the right-handed neutrino masses $m_N$.

The analysis of this section can be easily repeated for other flavour models based on different family symmetries and flavon alignment potentials to realise tribrid inflation in various different models that are motivated from particle physics.

\section{Inflationary trajectory without topological defects}
\label{sec:shiftedTribridA4}

As we have discussed in section \ref{sec:inflatonDirectionsTopological}, one can use different mechanisms to avoid the overproduction of topological defects at the end of inflation. In this section, we discuss a particular inflaton trajectory for which topological defects are avoided automatically, because the waterfall field has a small shift during inflation. In our model, this naturally occurs for inflation along the $LH_u$ direction with $\Theta_2$ as the waterfall field. We restrict our analysis to that special case throughout this section.

We will need to keep track of the different $A_4$ triplet components of both the inflaton and the waterfall field, so it will be useful to introduce the notation
\begin{align}
 \Theta_2 \, = \, \begin{pmatrix} x \\ y \\ z \end{pmatrix}, \quad LH_u \, = \, \Phi^2 \begin{pmatrix} 1 \\ 0 \\ 0 \end{pmatrix},
\end{align}
where we chose an inflaton direction which satisfies the condition $LH_u \perp \braket{ \Theta_1 }$ which we have explained in section \ref{sec:LHuDirections}.\footnote{This inflaton direction is not unique, one can choose any direction perpendicular to $\braket{\Theta_1}$. We focus on one specific direction, which suffices to demonstrate how the proposed mechanism works.}

We should also note that during inflation $\Theta_1$ can have any value
\begin{align}
 \Theta_1 \, = \, \begin{pmatrix} 0 \\ u \\ v \end{pmatrix},\quad \text{with }u^2+v^2 = \mu_1^2.
\end{align}
The flavon alignment potential \eqref{eq:flavonW} only fixes $u$ and $v$ when $\Theta_2$ has a sufficiently large vacuum expectation value, i.e.\ after the waterfall transition. During inflation, $u$ and $v$ are instead determined by their initial conditions.

\subsection{Shift in $\Theta_2$}

To see that $\Theta_2$ is non-zero during inflation, we need to minimize its scalar potential during inflation. The relevant superpotential terms are
\begin{align}
 W &\supset \lambda_2 \, \Theta_2 \cdot L H_u N_2  +  \gamma_{22} \braket{\Theta_S} N_2^2 \Theta_2^2  +  \kappa S_2 \left(  \Theta_2^3 - \mu_2^2  \right)   \notag\\
 &\quad +   \alpha_1 D_2'( \Theta_2^2 )_{1''}  +  \alpha_2 D_2''( \Theta_2^2 )_{1'} + \alpha_3 P_{12} \Theta_1 \Theta_2. \label{eq:flavon3W}
\end{align}
We use the $A_4$ contractions, with $\omega = e^{2\pi i / 3}$:
\begin{align}
 (\Theta_2^2)_{1'}  &=  x^2 + \omega^2 y^2 + \omega z^2, & \Theta_2^2  &=  x^2+y^2+z^2, \notag \\
 (\Theta_2^2)_{1''}  &=  x^2 + \omega y^2 + \omega^2 z^2, & \Theta_2^3  &=  xyz.
\end{align}
When calculating the F-term potential, we must include the K\"{a}hler potential coupling
\begin{align}
 K \, = \, (\kappa_{\Theta}+1) \lvert \Theta_2^2 S_2^2 \rvert + ...
\end{align}
The F-term potential then becomes
\begin{align}
V_F^{(\Theta_2)} \, &= \, \left| \frac{ \partial W }{ \partial N_2 } \right|^2  +  \left| \frac{ \partial W }{ \partial S_2 } \right|^2  +  \left| \frac{ \partial W }{ \partial D_2' } \right|^2  +  \left| \frac{ \partial W }{ \partial D_2'' } \right|^2   +  \left| \frac{ \partial W }{ \partial P_{12} } \right|^2  - \kappa_{\Theta} V_0 | \Theta_2 |^2 \notag \\
 &= \, \left|  \lambda_2 x \Phi^2 + 2\gamma_{22} \braket{\Theta_S} N_2 ( x^2 + y^2 + z^2 )  \right|^2  
 +  \kappa^2 \left|  xyz - \mu_2^2  \right|^2  
 - \kappa_{\Theta} V_0 ( |x|^2 + |y|^2 + |z|^2 )   \notag \\
 &\quad +  \alpha_1^2 \left|  x^2 + \omega y^2 + \omega^2 z^2  \right|^2  
 +  \alpha_2^2 \left| x^2 + \omega^2 y^2 + \omega z^2  \right|^2   +  \alpha_3^2 \left| \Theta_1 \cdot \Theta_2  \right|^2 .  \label{eq:Vxyz}
\end{align}

The last term enforces $\Theta_1 \perp \Theta_2$, which implies that
\begin{align}
 y = \varepsilon v, \quad z = -\varepsilon u, \quad \text{for some }\varepsilon \in \mathbb{C}  \label{eq:defineEpsilon}
\end{align}
and which also guarantees that we can ignore the operator $\gamma_{12} \braket{\Theta_S} \Theta_1 \Theta_2 N_1 N_2$ throughout this calculation.

\subsubsection{Shift in $y$ and $z$}
\label{sec:shiftyz}

To find the shifts in $y$ and $z$, we approximate $x \simeq 0$ and $N_2 \simeq 0$. We will later see under which conditions this approximation is valid. With this approximation and eq.~\eqref{eq:defineEpsilon}, the potential \eqref{eq:Vxyz} simplifies to
\begin{align}
 V_F^{(\Theta_2)}  \, &\simeq \,  V_0  -  \kappa_\Theta V_0 ( \lvert u \rvert^2 + \lvert v \rvert^2 ) \lvert \varepsilon \rvert^2  +  ( \alpha_1^2 + \alpha_2^2 )( \lvert u \rvert^4 + \lvert v \rvert^4 ) \lvert \varepsilon \rvert^4  \notag \\
 &\quad +  2 \, \lvert u v \rvert^2 \left\{  \alpha_1^2 \cos\left( 2 \arg(u/v) + \frac{2\pi}{3} \right)  +  \alpha_2^2 \cos\left( 2 \arg(u/v) - \frac{2\pi}{3} \right)  \right\} \lvert \varepsilon \rvert^4. \label{eq:VYZdelta}
\end{align}

Minimizing the potential with respect to $\lvert \varepsilon \rvert^2$ gives
\begin{align}
 \lvert \varepsilon \rvert^2 \, &\simeq \, \frac{ \lvert u \rvert^2 + \lvert v \rvert^2 }{ 2(\alpha_1^2 + \alpha_2^2)( \lvert u \rvert^4 + \lvert v \rvert^4 ) + 4 \gamma \lvert uv \rvert^2  } \, \kappa_\Theta V_0, \label{eq:epsilonSol}
\end{align}
with
\begin{align}
 \gamma \, :=& \, \alpha_1^2 \cos\left( 2 \arg(u/v) + \frac{2\pi}{3} \right)  +  \alpha_2^2 \cos\left( 2 \arg(u/v) - \frac{2\pi}{3} \right).
\end{align}
For $\alpha_i = O(1)$, one can show that eq.~\eqref{eq:epsilonSol} implies $\operatorname{max}(\lvert y \rvert, \lvert z \rvert) = O(1) \sqrt{\kappa_\Theta V_0}$.

\emptyline

Note that the potential is not minimized with respect to $\Theta_1 = (0, u, v)$. Instead, $u$ and $v$ are constants given by the initial conditions, because in general their potential is very flat compared to the inflaton potential, and therefore the fields can be treated as constant during inflation.\footnote{$u$ and $v$ initially minimize the F-term potential for $\Theta_2 = 0$, so the only term which induces a potential for $u$ and $v$ is the interaction term $\alpha_3^2 \lvert \Theta_1 \cdot \Theta_2 \rvert^2$, which creates a potential for $\Theta_1$ and $\Theta_2$ when they are not perpendicular to each other. However, this potential for each field is proportional to the vacuum expectation value of the other field, and due to $\Theta_1 \sim \mu_1 \gg \Theta_2 \sim \Hubble$, the potential will generally be much greater for $\Theta_2$ and thus be minimized by an adjustment in $\Theta_2$.}

\subsubsection{Shift in $x$}
Using the shifts in $y$ and $z$, but still assuming that $S_2 = N_2 = 0$, we can find the minimum for $x$ by minimizing eq.~\eqref{eq:Vxyz}:
\begin{align}
 V_F^{(x)} \, &= \, (\lambda_2^2 \lvert \Phi \rvert^4 - \kappa_\Theta V_0)  \lvert x \rvert^2
 -  2 \kappa^2 \mu_2^2 \operatorname{Re}(xyz) + (\alpha_1^2 + \alpha_2^2) \vert x \rvert^4  \notag \\
 &\quad  + 2 \alpha_1^2 \operatorname{Re}\left\{ (x^*)^2 (\omega y^2 + \omega^2 z^2 ) \right\}
 + 2 \alpha_2^2 \operatorname{Re}\left\{ (x^*)^2 (\omega^2 y^2 + \omega z^2 ) \right\}.
 \label{eq:Vxyz2}
\end{align}
The two dominant terms are $\lambda_2^2 \lvert \Phi \rvert^4$ and the linear term in $x$. The other mass terms for $x$ are negligible during inflation because $\lambda_2^2 \lvert \Phi \rvert^4 \gg \lambda_2^2 \lvert \Phi_c \rvert^4 = \kappa_\Theta V_0$, and $y^2$, $z^2 \lesssim \kappa_\Theta V_0$. The $x^4$ term is also negligible, as we will find that $x^2 \ll \kappa_\Theta V_0$.

We can therefore minimize the simple potential
\begin{align}
V_F^{(x)} \, &\simeq \, \lambda_2^2 \lvert \Phi \rvert^4  \lvert x \rvert^2  -  2 \kappa^2 \mu_2^2 \operatorname{Re}(xyz)
\end{align}
with respect to $x$. The result is
\begin{align}
\lvert x \rvert \, &\simeq \, \frac{ \kappa \lvert yz \rvert }{ \lambda_2^2 \lvert \Phi \rvert^4 } \underbrace{ \kappa \mu_2^2 }_{\sqrt{V_0}} \, \leq \, \frac{ \kappa \operatorname{max}(\lvert y \rvert, \lvert z \rvert)^2 }{ \lambda_2^2 \lvert \Phi \rvert^4 } \sqrt{V_0} \, = \, O(1) \, \kappa  \frac{ \kappa_\Theta V_0 }{ \lambda_2^2 \lvert \Phi \rvert^4 } \sqrt{V_0}.
\end{align}
During inflation, we have $\lambda_2^2 \Phi^4 \gg \lambda_2^2 \Phi_c^4 = \kappa_\Theta V_0$, and therefore $\lvert x \rvert^2 \ll V_0$, so $x$ is suppressed with respect to $y$, $z \sim \sqrt{\kappa_\Theta V_0}$. This justifies our assumption in section~\ref{sec:shiftyz}, where we set $x \simeq 0$ for the calculation of the shifts in $y$ and $z$.

Near the critical point, we find that $x$ is no longer suppressed with respect to $y$ and $z$: $x \simeq O(1) \, \sqrt{V_0}$. In this case, $y$ and $z$ are no longer stabilized by $|\partial W / \partial D_2'|^2$ and $|\partial W / \partial D_2''|^2$, and we expect $\Theta_2$ to roll down the potential along the direction $x \simeq y \simeq z$. Therefore, near the critical inflaton value, we could have a smooth transition towards the minimum of the waterfall field instead of a sudden waterfall at $\Phi = \Phi_c$. Anyway, as the waterfall potential is steep, we expect inflation to end quickly near $\Phi \simeq \Phi_c$, so that the model still approximately acts as a model of K\"{a}hler-driven tribrid inflation.

We expect that the predictions for $\alpha_s$ and $\braket{H}$ are not influenced by the shift in $\Theta_2$, as they depend only on the inflaton potential far away from the critical point. However, $\lambda_2$ (and the derived $y_{\nu_2}$) might be changed by approximately up to a factor of $\sim \exp(\pm \delta N / 10)$, where $\delta N$ is the number of e-folds before the critical point at which the shift in $\Theta_2$ becomes relevant.\footnote{This change in $\lambda_2$ happens if the shift in $\Theta_2$ either accelerates or delays the phase transition, because $\lambda_2$ must be adjusted to get the correct number of e-folds between $\Phi_0$ and the end of inflation. A precise prediction of $\lambda_2$ therefore requires an analysis of the complete multi-field model near $\Phi \simeq \Phi_c$ including all components of $\Theta_2$, and possibly a quantitative treatment of the tachyonic preheating phase.}

\subsubsection{Shifts in other fields}

$S_2$ and $N_2$ also get small shifts due to the shift in $x$. However, these are completely negligible if $S_2$ and $N_2$ also get positive Hubble-size masses from the K\"{a}hler potential: in that case, one can easily show that $S_2 \ll N_2 \ll \lvert x \rvert$. We therefore set $S_2$, $N_2 \simeq 0$ in this paper.

\subsection{Inflaton potential}

The shift in $\Theta_2$ induces a small correction to the inflaton potential from
\begin{align}
 V_F \, &\supset \, \left|  \lambda_2 \Theta_2 \cdot L H_u  \right|^2 \,
 = \, \lambda_2^2 \lvert x \rvert^2 \left| \Phi \right|^4 \, \lesssim \, \lambda_2^2 \kappa^2 V_0   \left(  \frac{ \kappa_\Theta V_0 }{ \lambda_2^2 \lvert \Phi \rvert^4 } \right)^2  \left| \Phi \right|^4 \notag \\
 &\ll \, O(1)\,V_0 \left|\Phi \right|^4, \quad (\text{for }\Phi \gg \Phi_c).
\end{align}
For $\Phi \gg \Phi_c$, the extra inflaton potential that is induced by $\Theta_2 \neq 0$ is therefore much smaller than the inflaton potential $V_\phi \simeq V_0(1 + a \, \phi^2 + b \, \phi^4)$.

We conclude that despite the shift in $\Theta_2$, inflation is mostly K\"{a}hler-driven, except possibly near $\Phi \simeq \Phi_c$ around which we expect inflation to end anyway. Therefore we can approximately use the predictions for $y_{\nu_2} \, \hat{=} \, y_\nu$ and $\braket{\Theta_2} \, \hat{=} \, \braket{H}$ from fig.~\ref{fig:l3predict}, which have been calculated for purely K\"{a}hler-driven tribrid inflation.

\subsection{Possible variant: $S_2( \Theta_2^5 - \mu_2^2 )$}

A simple variant of this model can be obtained by replacing the vacuum energy term for $\Theta_2$ in the superpotential:
\begin{align}
 W \, \supset \, \kappa S_2( \Theta_2^3 - \mu_2^2 )  ~ \rightarrow ~ \kappa S_2\left( \Theta_2^3\, \Theta_2^2 - \mu_2^2 \right).
\end{align}
The replacement suppresses the shift in $x$ during inflation by another power of $y^2 + z^2 \sim V_0 \sim 10^{-12}$, which makes it completely negligible. This improves the quality of our approximations and we expect that the model will behave more exactly like purely K\"{a}hler-driven tribrid inflation, even very close to $\Phi \simeq \Phi_c$. Nevertheless, this setup may still remove domain walls during the waterfall transition, as the potential still prefers a specific phase for $x$ when $x$, $y$ and $z$ grow during the waterfall transition.

However, it is not clear whether the preference is sufficiently large to either prevent the production of domain walls or to efficiently remove them if they are produced. To answer this question, the preheating phase should be studied in detail, which is beyond the scope of this paper.

\subsection{Conditions for applying this mechanism}

\label{sec:topologicalMechanism}

We have seen that if the flavon $\Theta_2$ is identified with the waterfall field and $LH_u$ with the inflaton, the waterfall field obtains a small shift during inflation, which can help with preventing domain walls. However, if we had used $\Theta_1$ or $\Theta_3$ as the waterfall field or $N_2$ as the inflaton, no such shift would have occured during inflation.

In this section, we briefly want to sketch which characteristics of the superpotential determine whether the waterfall field obtains a shift during K\"{a}hler-driven tribrid inflation or not.

The main condition can be split in three parts:
\begin{enumerate}
 \item The waterfall field $H$ consists of several fields, e.g.\ it is a multiplet under some symmetry.
 \item At least one of the multiplet components of $H$ does not receive a mass from the inflaton.
 \item The same multiplet component of $H$ does not receive a mass term from the alignment potential, but is instead stabilized by higher-order terms only.\footnote{There must be stabilizing terms, otherwise the waterfall will happen immediately along the unstabilized direction.}
\end{enumerate}

Let us apply these considerations to our example model: The first condition restricts the range of waterfall fields for which this mechanism can work. In our model it is satisfied because we use an $A_4$ triplet flavon as the waterfall field.
The second condition explains why we do not get a shifted waterfall field when we use $N_2$ as the inflaton: the superpotential term $W_N = \gamma_{22} \braket{\Theta_S} \Theta_2^2 N_2^2 = \gamma_{22} \braket{\Theta_S} (x^2 + y^2 + z^2)N_2^2$ provides masses to all three $A_4$ components of $\Theta_2$ as long as $N_2 > \Phi_c$, so the waterfall field is stabilized exactly at zero during inflation.
The third condition explains why using $\Theta_1$ or $\Theta_3$ does not lead to shifted waterfall trajectories. Although an $LH_u$ inflaton only provides masses to one of the three $A_4$ components of $\Theta_j$, the other two components acquire mass terms from the flavon alignment potential via the terms $W_{\text{fl}}^{(\Theta_1)} \supset P_{12} \Theta_1 \Theta_2  +  P_{16} \Theta_1 \Theta_6$ and $W_{\text{fl}}^{(\Theta_3)} \supset P_{34} \Theta_3 \Theta_4  +  P_{36} \Theta_3 \Theta_6$. Therefore, during inflation all three $A_4$ directions of the waterfall field are stabilized at zero, and no shift can occur.

\section{Summary and conclusions}

We have discussed how K\"{a}hler-driven tribrid inflation can be implemented in realistic particle physics models, using a D-flat combination of charged matter fields as the inflaton and a Higgs field as the waterfall field to terminate inflation. We explained how one can identify suitable inflaton and waterfall fields and which predictions tribrid inflation makes for the model parameters. Our strategy can be easily applied to a wide range of models based on different symmetries, leading to predictive models of inflation in motivated particle physics models.

As an explicit example, we have analysed a particular flavour model based on a spontaneously broken $A_4$ family symmetry. The inflaton can be either a right-handed sneutrino or a D-flat $LH_u$ direction, and tribrid inflation predicts a relation between the running of the spectral index $\alpha_s$ and either the right-handed neutrino mass $m_N$ or the neutrino Yukawa coupling $y_\nu$. The model also predicts $\theta_{23} = 45^\circ$ and a relation between the mixing angles and the leptonic Dirac CP phase.

Further predictions could be obtained from a detailed study of the reheating phase \cite{reheating}. As the inflaton is composed of MSSM fields, its couplings to the visible sector are known (or could be measured with a collider), so one could calculate extra bounds from leptogenesis or from possible overproduction of gravitinos or dark matter. Though such an analysis is beyond the scope of our paper, it is important to realize that by realizing inflation in the visible matter sector, models of tribrid inflation can be very predictive in principle.

We have also shown that in our $A_4$ model, the formation of topological defects at the end of inflation can be avoided automatically, because the waterfall field can have a small shift. This is related to our choice of the flavon alignment potential. We expect that such a shift could be generated in models with other discrete family symmetries as well.

\section*{Acknowledgements}
This work was supported by the Swiss National Science Foundation. We thank Vinzenz Maurer and Constantin Sluka for useful discussions.

\section*{Appendix}
\appendix

\section{Predictions for K\"{a}hler-driven tribrid inflation}
\label{appendix:kahlerplots}
In this section, we provide plots for the predictions for K\"{a}hler-driven tribrid inflation for different superpotential parameters $\ell$. The plots are based on the calculations in \cite{kahlerdriven} and normalized to the experimental data from the Planck satellite \cite{Planck}.

\begin{figure}[tbhp]
  \centering$
\begin{array}{cc}
\includegraphics[width=0.48\textwidth]{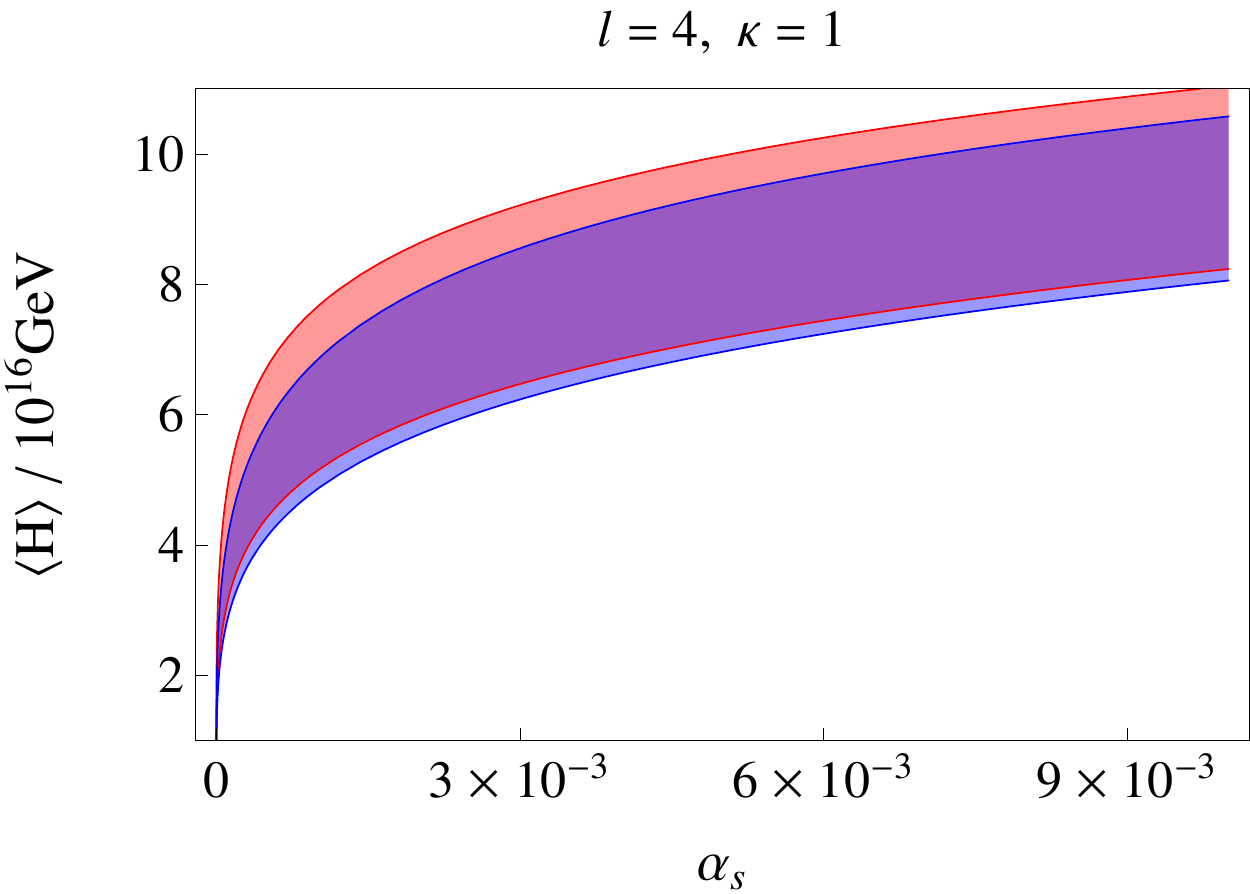} &
\includegraphics[width=0.48\textwidth]{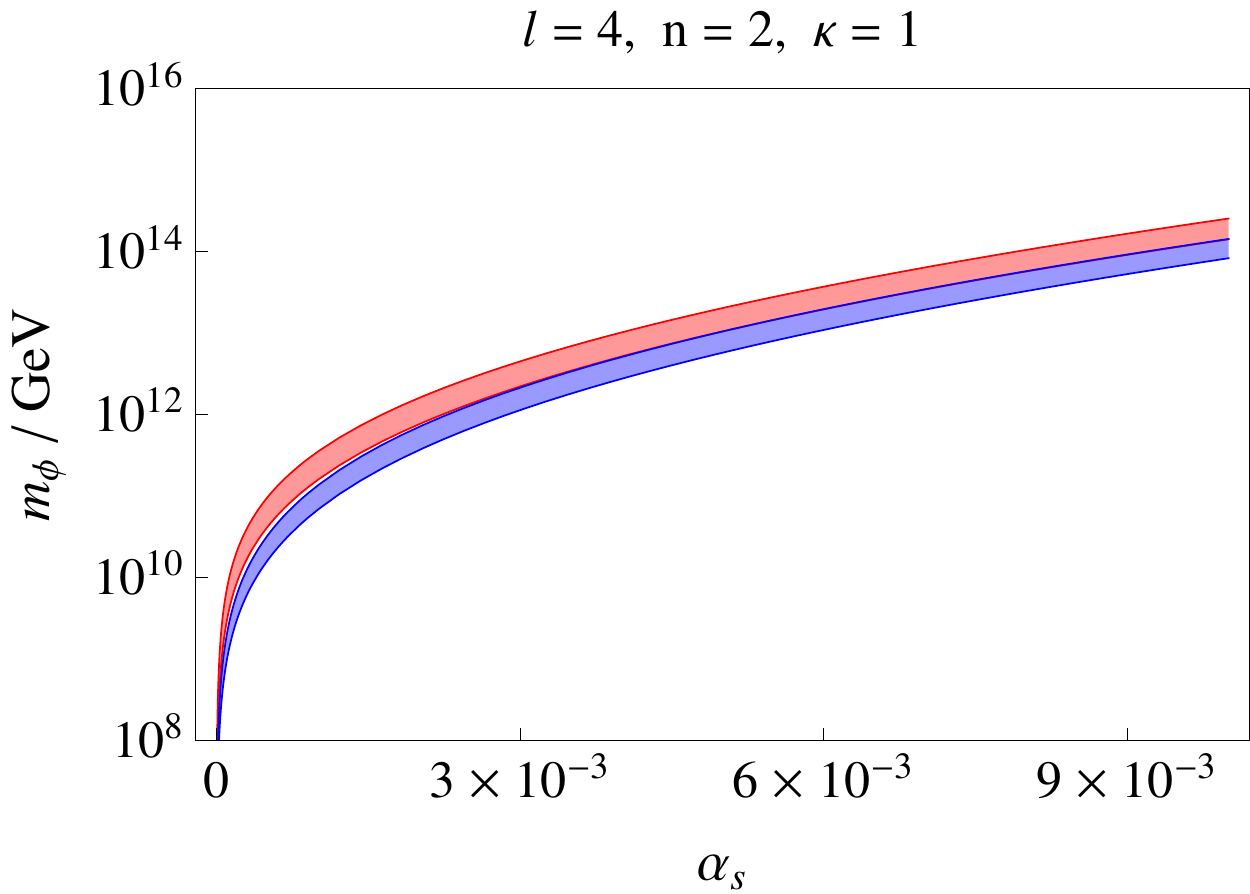} \\
\includegraphics[width=0.48\textwidth]{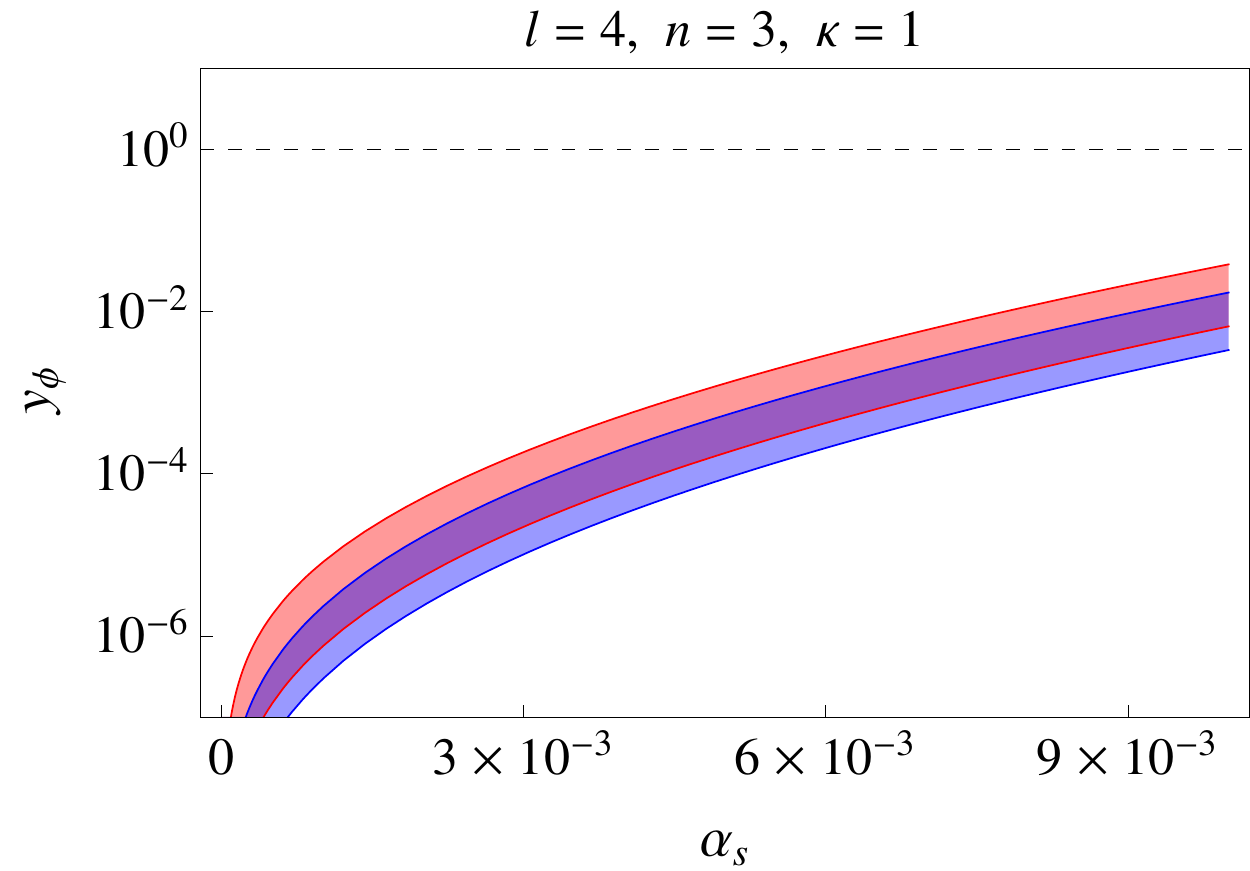} &
\includegraphics[width=0.48\textwidth]{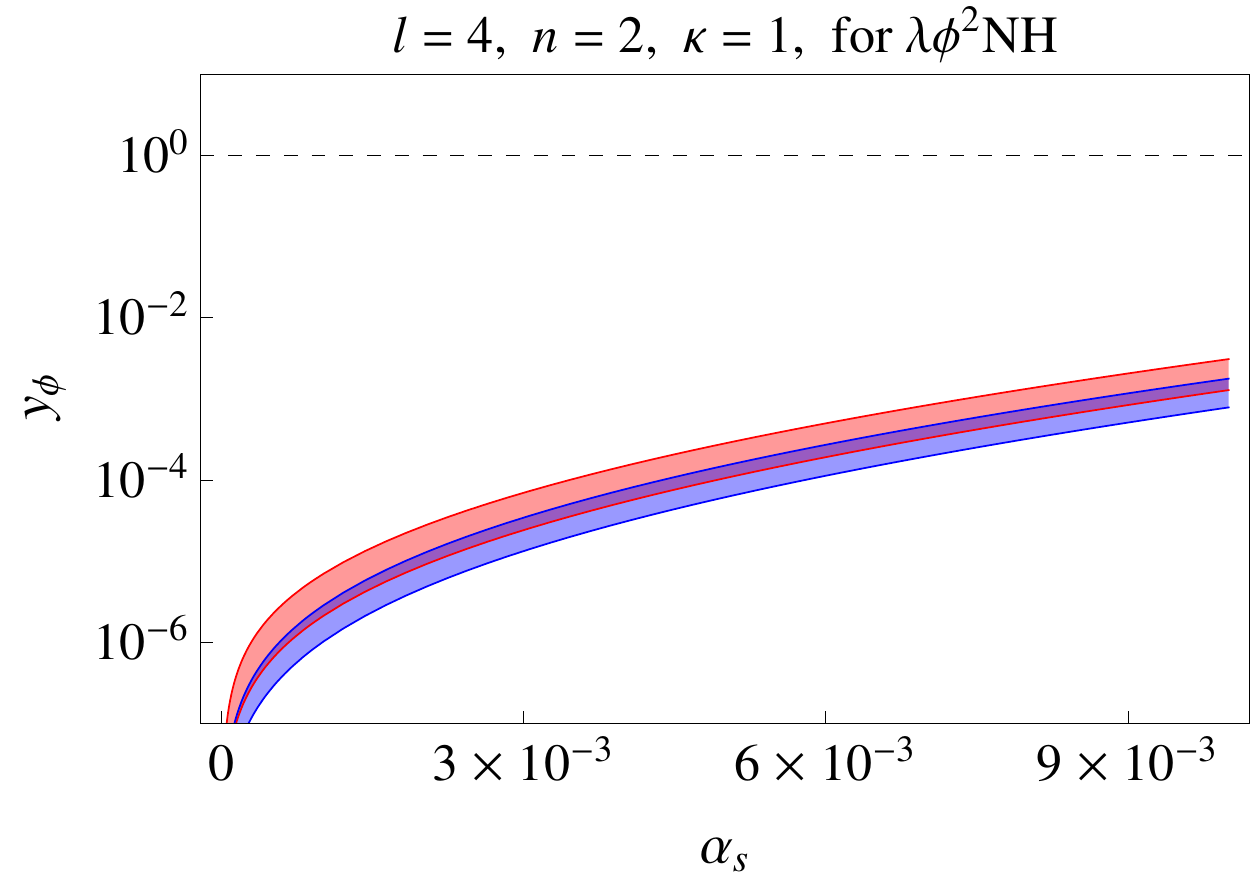}
\end{array}$
  \caption{Predicted values for the waterfall vacuum expectation value $\braket{H}$ and the inflaton mass $m_\Phi$ or Yukawa coupling $y_\Phi$, depending on the running of the spectral index $\alpha_s$. Results are shown for $n_s = 0.946$ (blue band) and $n_s = 0.975$ (red band), $N_0 = 55$, $\ell=4$, $\kappa = 1$ and $\kappa_{011} = 2$. For other values of $\kappa$ and $\kappa_{011}$, these predictions scale with $\braket{H} \propto \kappa^{-1/4}$ and $\lambda \propto (\kappa_{011} - 1)^{1/2}$. The parameters $\ell,n,\lambda$ and $\kappa$ are defined in eq.~(\ref{eq:tribridW}), and $\kappa_{011}$ is a parameter of the K\"ahler potential as defined in ref.~\cite{kahlerdriven}. The width of the bands is due to free parameters of the K\"{a}hler potential related to the quartic inflaton coupling \cite{kahlerdriven}.}
  \label{fig:l4predict}
\end{figure}

For $\ell \geq 3$, loop corrections are generally suppressed, and the results should be fair approximations to the exact slow-roll results. More details can be found in \cite{kahlerdriven}.

For $\ell=2$, tribrid inflation can be successfully realized, but the predictions depend on both the K\"{a}hler potential and the one-loop effective potential. In some cases, the loop potential dominates, e.g.\ if the K\"{a}hler potential contributions vanish due to some symmetry \cite{loopdriven1,tribridHeisenberg}; in other cases, the K\"{a}hler potential is dominant and the loop potential can be neglected. In the more general mixed case, however, one should include both contributions to the inflaton potential for calculating the slow-roll predictions.

\begin{figure}[tbhp]
  \centering$
\begin{array}{cc}
\includegraphics[width=0.48\textwidth]{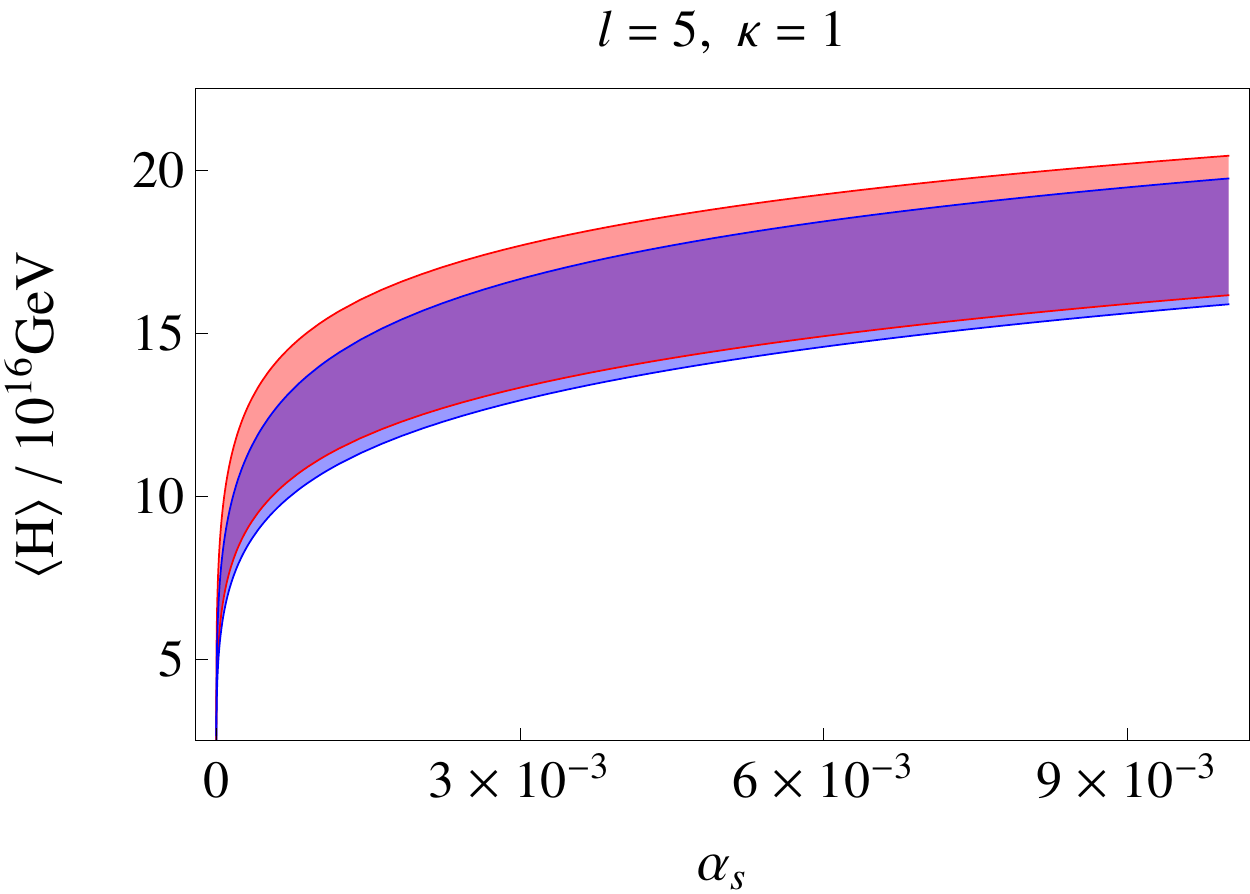} &
\includegraphics[width=0.48\textwidth]{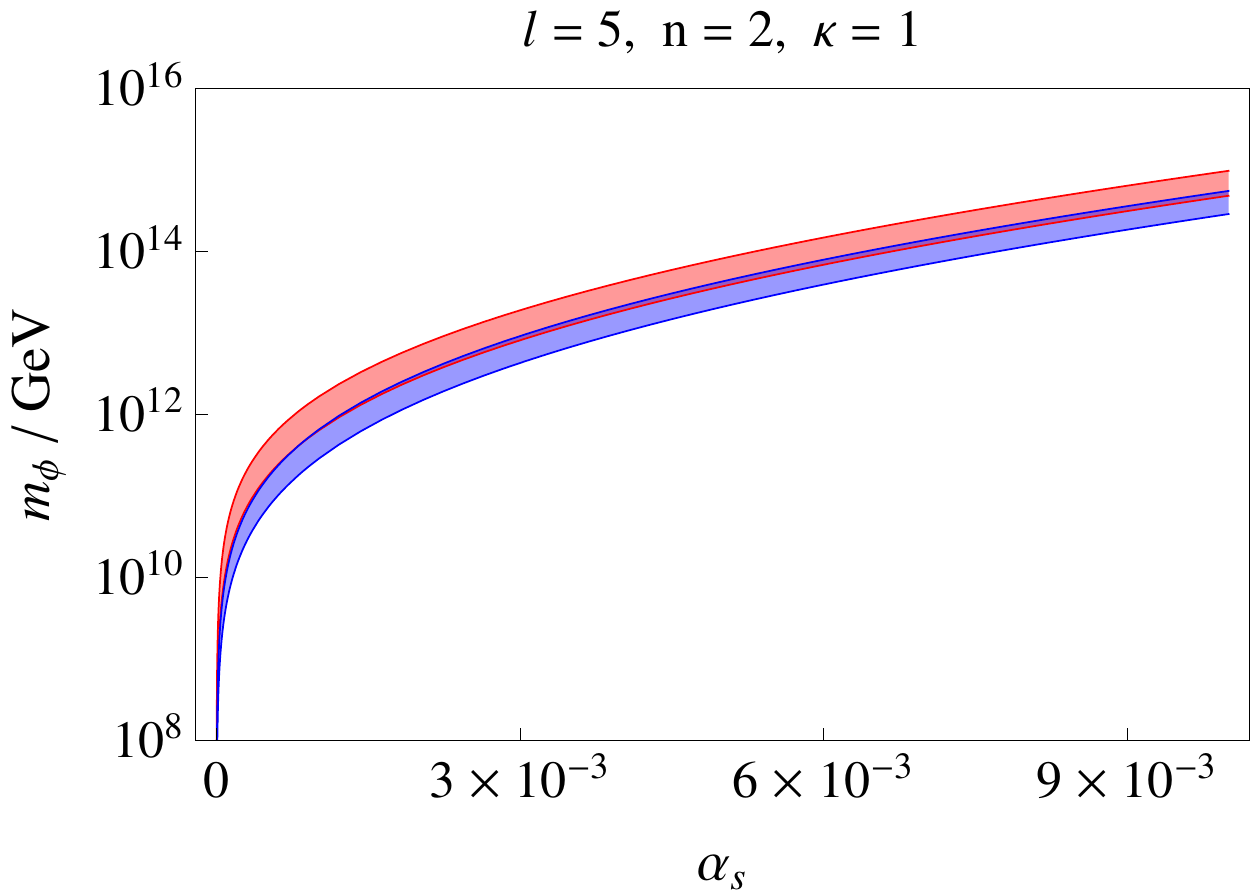} \\
\includegraphics[width=0.48\textwidth]{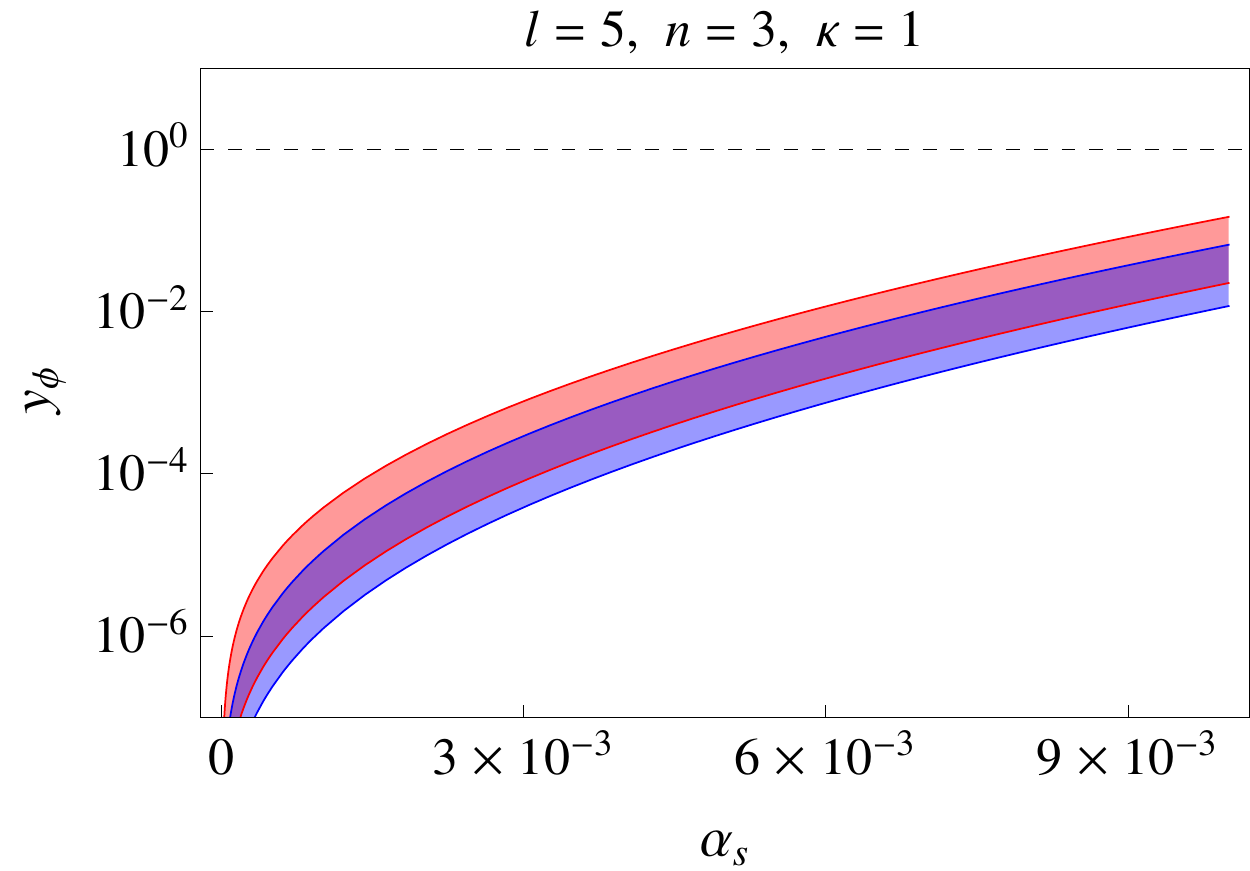} &
\includegraphics[width=0.48\textwidth]{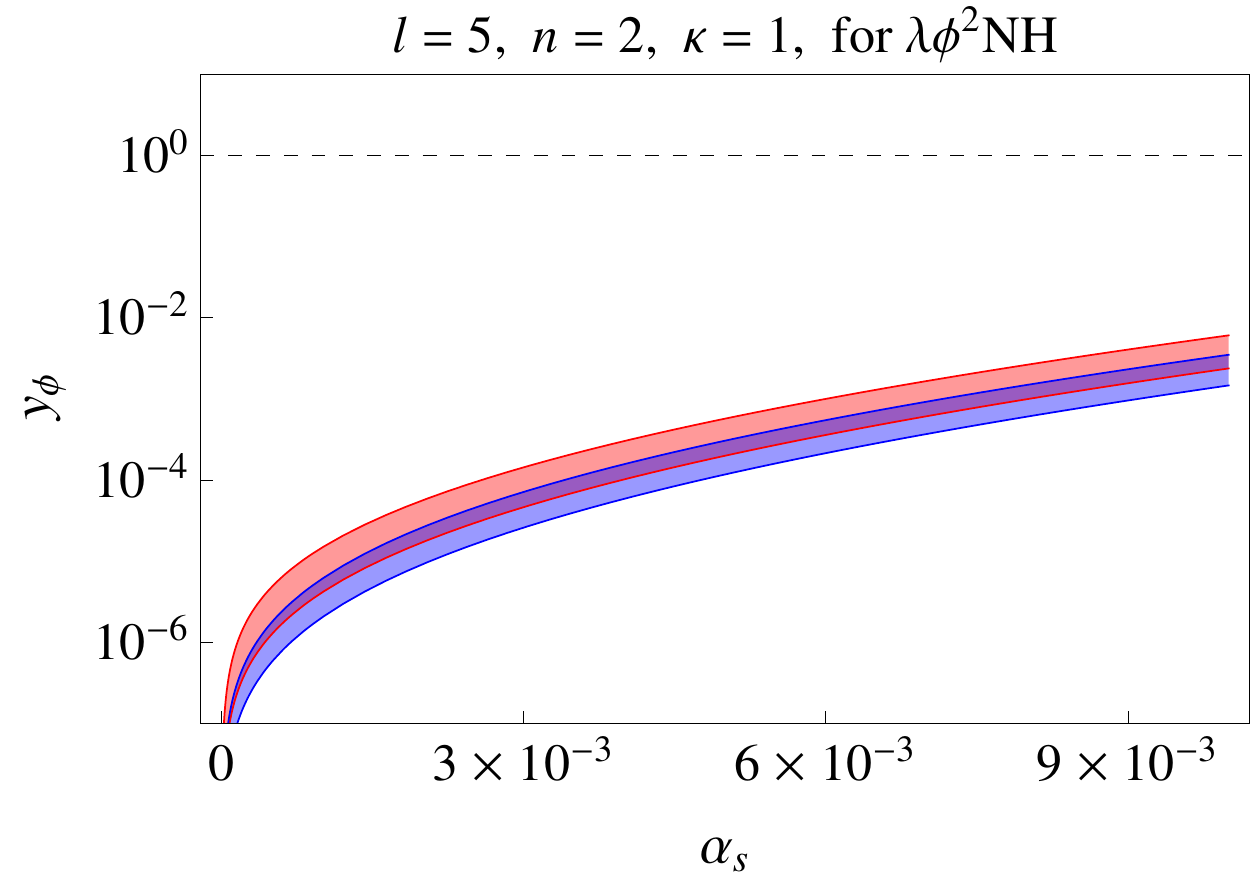}
\end{array}$
  \caption{Predicted values for the waterfall vacuum expectation value $\braket{H}$ and the inflaton mass $m_\Phi$ or Yukawa coupling $y_\Phi$, depending on the running of the spectral index $\alpha_s$. Results are shown for $n_s = 0.946$ (blue band) and $n_s = 0.975$ (red band), $N_0 = 55$, $\ell=5$, $\kappa = 1$ and $\kappa_{011} = 2$. For other values of $\kappa$ and $\kappa_{011}$, these predictions scale with $\braket{H} \propto \kappa^{-1/5}$ and $\lambda \propto (\kappa_{011} - 1)^{1/2}$. The parameters $\ell,n,\lambda$ and $\kappa$ are defined in eq.~(\ref{eq:tribridW}), and $\kappa_{011}$ is a parameter of the K\"ahler potential as defined in ref.~\cite{kahlerdriven}. The width of the bands is due to free parameters of the K\"{a}hler potential related to the quartic inflaton coupling \cite{kahlerdriven}.}
  \label{fig:l5predict}
\end{figure}

\begin{figure}[tbhp]
  \centering$
\begin{array}{cc}
\includegraphics[width=0.48\textwidth]{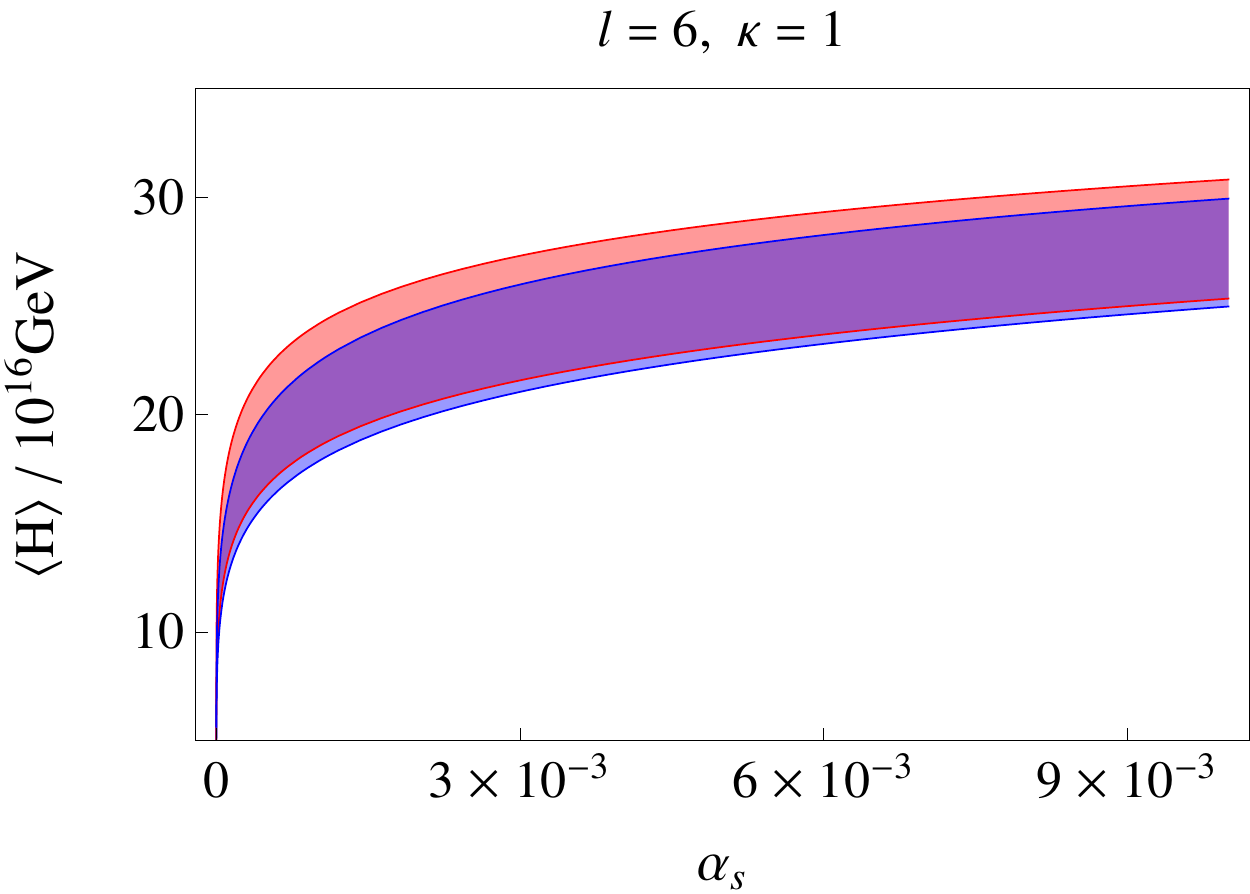} &
\includegraphics[width=0.48\textwidth]{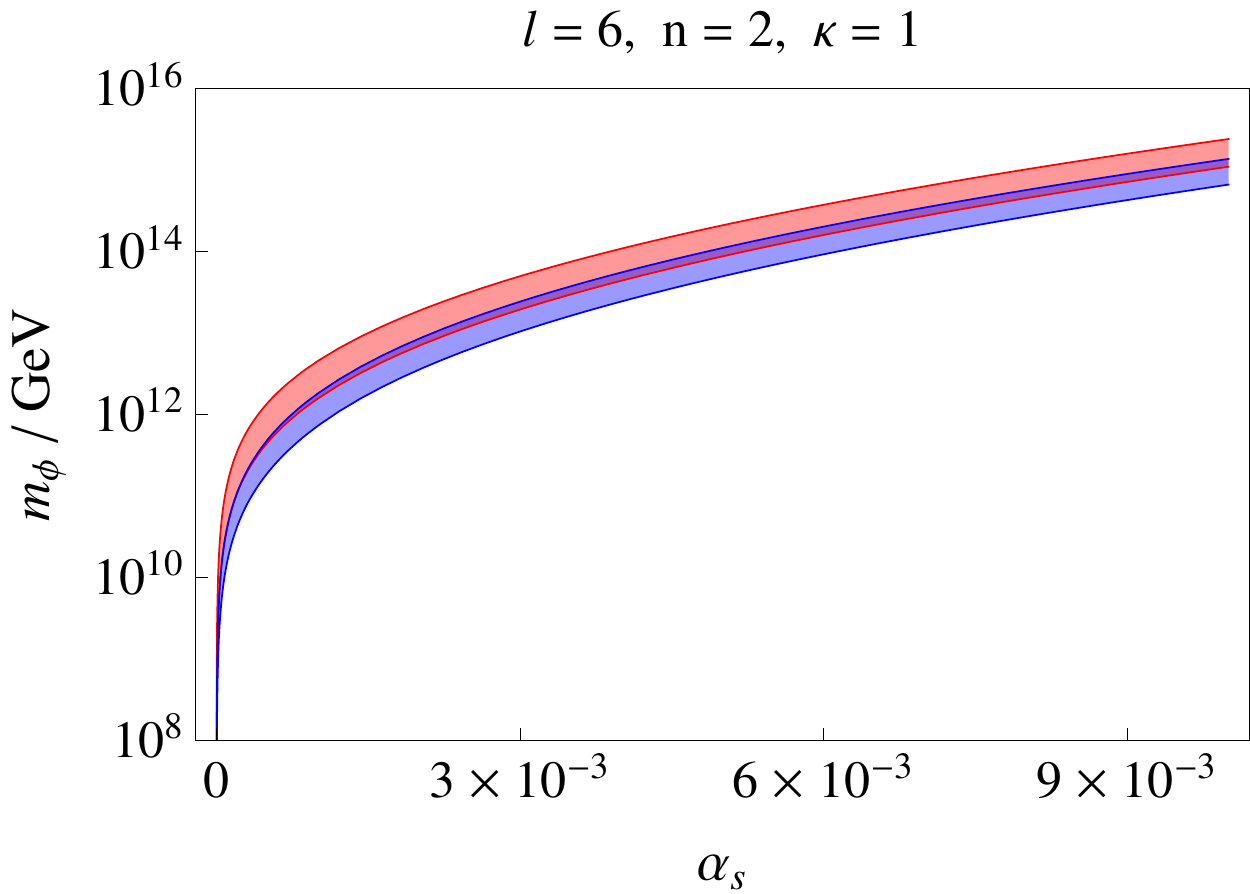} \\
\includegraphics[width=0.48\textwidth]{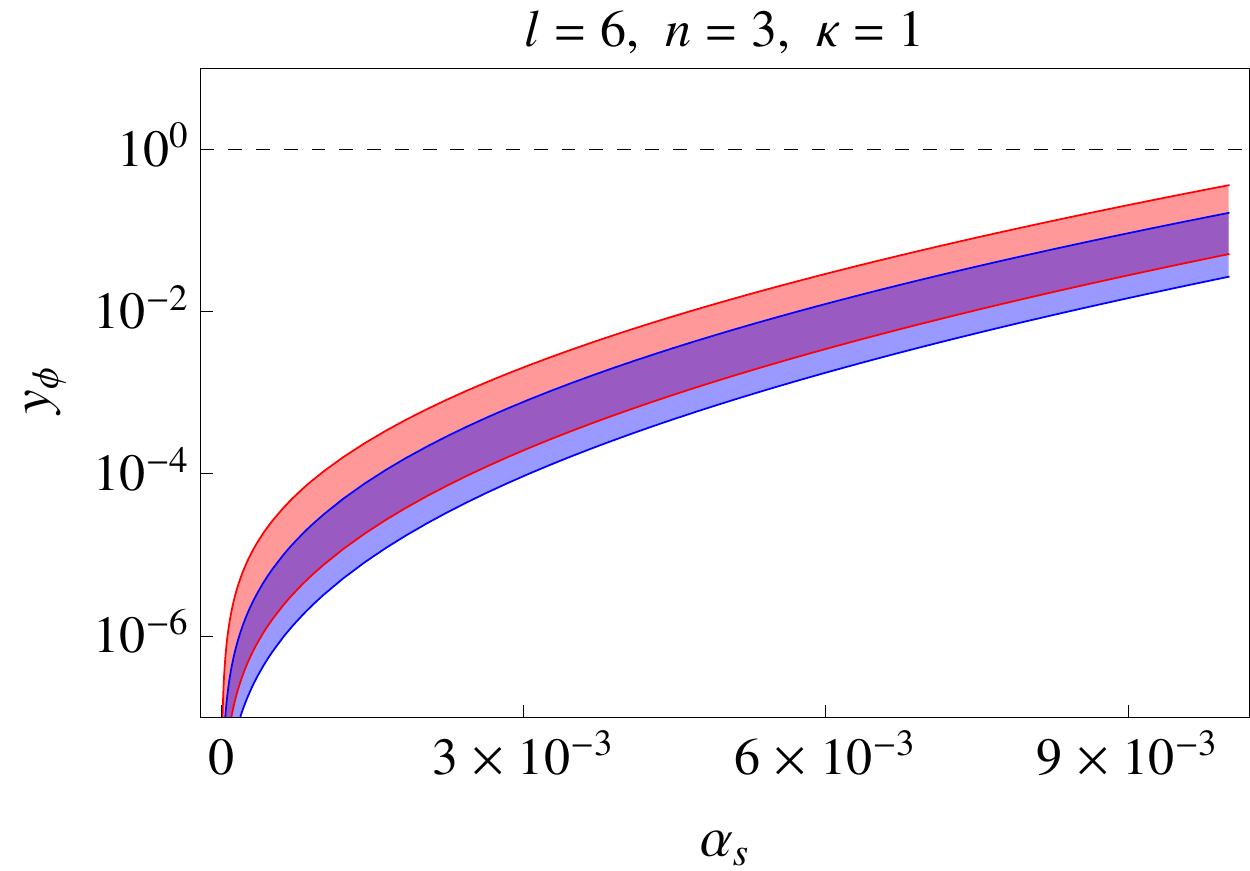} &
\includegraphics[width=0.48\textwidth]{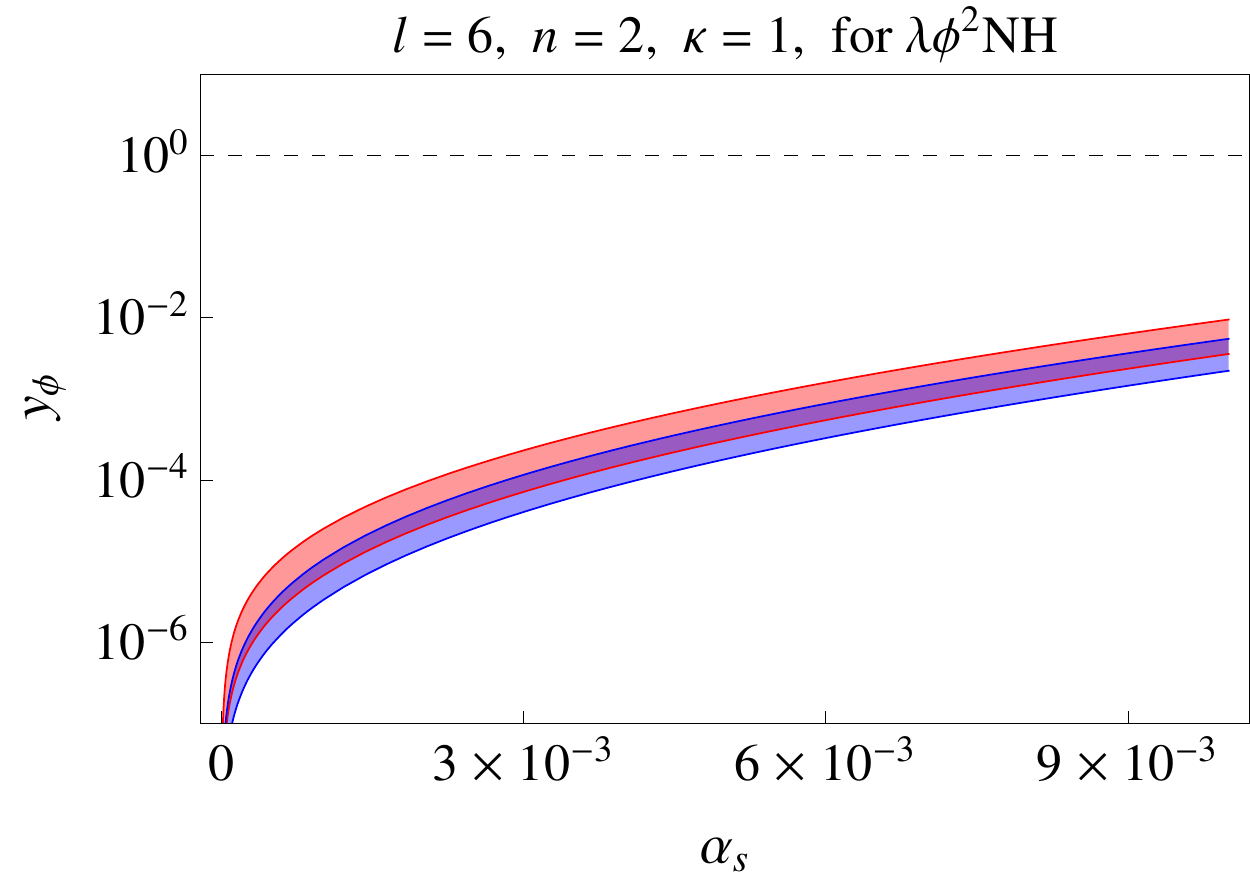}
\end{array}$
  \caption{Predicted values for the waterfall vacuum expectation value $\braket{H}$ and the inflaton mass $m_\Phi$ or Yukawa coupling $y_\Phi$, depending on the running of the spectral index $\alpha_s$. Results are shown for $n_s = 0.946$ (blue band) and $n_s = 0.975$ (red band), $N_0 = 55$, $\ell=6$, $\kappa = 1$ and $\kappa_{011} = 2$. For other values of $\kappa$ and $\kappa_{011}$, these predictions scale with $\braket{H} \propto \kappa^{-1/6}$ and $\lambda \propto (\kappa_{011} - 1)^{1/2}$. The parameters $\ell,n,\lambda$ and $\kappa$ are defined in eq.~(\ref{eq:tribridW}), and $\kappa_{011}$ is a parameter of the K\"ahler potential as defined in ref.~\cite{kahlerdriven}. The width of the bands is due to free parameters of the K\"{a}hler potential related to the quartic inflaton coupling \cite{kahlerdriven}.}
  \label{fig:l6predict}
\end{figure}

\section{K\"{a}hler-driven tribrid inflation potential for composite inflaton}
When we compose the inflaton as $\Phi^n = \Phi_1 \Phi_2 \cdots \Phi_n$, e.g.\ $\Phi^2 := L H_u$, the inflaton potential $V(\phi_1, \phi_2, ..., \phi_n)$ depends on all $\phi_i$ independently. In this section, we show that the inflaton trajectory during K\"{a}hler-driven tribrid inflation nevertheless has the same inflaton potential as if it was a single scalar field. This ensures that the predictions for the single-field case still apply even when we replace $\Phi^n \rightarrow \Phi_1 \Phi_2 \cdots \Phi_n$.\footnote{The results are influenced only by a numerical factor when taking derivatives with respect to $\Phi$, which shifts $\lambda$ by a factor of $O(1)$.}

\subsection{Multifield inflaton potential}

We assume that the K\"{a}hler potential depends only on the modulus squared of the superfields:
\begin{align}
 K  =  \lvert H \rvert^2 + \lvert S \rvert^2 + \sum_i \left( \lvert \Phi_i \rvert^2  +  \kappa_{iS} \lvert \Phi_i S \rvert^2 \right)
   + \sum_{i \leq j} \left( \kappa_{i j} \lvert \Phi_i \Phi_j \rvert^2  +  \kappa_{i j S} \lvert \Phi_i \Phi_j S \rvert^2 \right) + ...
\end{align}

The inflaton potential, expressed in terms of canonically normalized real scalar fields $\phi_i = \frac{1}{\sqrt{2}} \lvert \Phi_i \rvert$, then turns out to be:
\begin{align}
 V(\phi_1, ..., \phi_n) \, &= \, V_0 \left( 1 + \sum\limits_i a_i \phi_i^2 + \sum\limits_{i \leq j} b_{ij} \phi_i^2 \phi_j^2 + ... \right) \, + \, V_D,
\end{align}
with the coefficients\footnote{We neglect effects from canonical normalization. For tribrid inflation with a single inflaton field, it has been shown explicitly that these effects can be interpreted as higher-order corrections to the potential, which can be absorbed e.g.\ in the definition of $b$. \cite{kahlerdriven}}
\begin{subequations}
\begin{align}
 a_i \, &= \, \frac{1}{2} \left(  1 - \kappa_{i S}  \right), \\
 b_{ii} \, &= \, \frac{1}{4} \left( \frac{1}{2} + \kappa_{ i i } - \kappa_{ i i S } +  \kappa_{i S}^2  -  \kappa_{i S}   \right) \, := \, b_i, \\
 b_{ij} \, &= \, \frac{1}{4} \left( 1 + \kappa_{ i j } - \kappa_{ i j S } +  2\kappa_{i S}\kappa_{j S}  -  \kappa_{i S}  -  \kappa_{j S}   \right), \quad (\text{for }i < j).
\end{align}
\end{subequations}
These coefficients depend entirely on the K\"{a}hler potential, and if the cutoff scale is around the Planck scale, we find that $a_i$, $b_{ij} \lesssim O(1)$.

\subsection{Deviation from D-flatness}

In this section, we work with only two fields $\Phi_1 = L$ and $\Phi_2 = H_u$ to keep the discussion simple:
\begin{align}
 V(\phi_1, \phi_2) \, &= \, V_0 \left( 1 + a_1 \, \phi_1^2 + a_2 \, \phi_2^2 + b_{1} \, \phi_1^4  +  b_{2} \, \phi_2^4   +  b_{12} \, \phi_1^2 \phi_2^2 \right) \, + \, \frac{1}{2}\sum\limits_{a}g_a^2 \left( \Phi^\dagger T^a \Phi \right)^2.
\end{align}
From this function, we want to derive the one-dimensional inflaton potential $V(\phi)$ along the inflaton trajectory.

The D-term will generally depend on the gauge charges of the $\phi_i$. In our example $\Phi_1 = L$ and $\Phi_2 = H_u$:
\begin{align}
 V_D \, &= \, \frac{g_1^2}{8} \left[ \left( L^\dagger, H_u^\dagger \right) \begin{pmatrix} -1 & 0 \\ 0 & 1 \end{pmatrix} \begin{pmatrix} L \\ H_u \end{pmatrix} \right]^2 
 + \frac{g_2^2}{8} \sum\limits_{i=1}^{3} \left[ \left( L^\dagger, H_u^\dagger \right) \begin{pmatrix} \sigma^i & 0 \\ 0 & \sigma^i \end{pmatrix} \begin{pmatrix} L \\ H_u \end{pmatrix} \right]^2 \notag \\
 &= \, \frac{g_1^2}{8} \left( \lvert L \rvert^2 - \lvert H_u \rvert^2 \right)^2 + \frac{g_2^2}{8} \sum\limits_{i=1}^{3} \left( L^\dagger \sigma^i L + H_u^\dagger \sigma^i H_u \right)^2.
\end{align}
The second term enforces that $L$ and $H_u$ have opposite weak isospin. As the inflaton potential does not otherwise depend on the $SU(2)_L$ structure of $L$ and $H_u$, this condition will be satisfied during inflation, and the term will be zero.

The first term enters the inflaton potential $V(\phi_1, \phi_2)$:
\begin{align}
 V(\phi_1, \phi_2) \, &= \, V_0 \left( 1 + a_1 \, \phi_1^2 + a_2 \, \phi_2^2 + b_{1} \, \phi_1^4  +  b_{2} \, \phi_2^4   +  b_{12} \, \phi_1^2 \phi_2^2 \right) \, + \, \frac{g_1^2}{32} \left( \phi_1^2 - \phi_2^2 \right)^2 \notag \\
 &= \, V_0 \left\{ 1 + a_1 \, \phi_1^2 + a_2 \, \phi_2^2 + ( b_{1} + d ) \phi_1^4  +  ( b_{2} + d ) \, \phi_2^4   +  ( b_{12} - 2d ) \, \phi_1^2 \phi_2^2 \right\}, \label{eq:Vphi1phi2}
\end{align}
where we defined $d := \frac{g_1^2}{32V_0} \sim O(10^{10})$. The terms with $d$ dominate over the small coefficients $a_{i}$, $b_{ij} \lesssim O(1)$. Therefore, we expect that the inflaton trajectory will be along a nearly D-flat direction.

Looking at trajectories near the D-flat trajectory $\phi_1^2 = \phi_2^2$, we define
\begin{align}
 \phi_1 = \frac{1}{\sqrt{2}} \left( \phi + \delta \right), \quad \phi_2 =  \frac{1}{\sqrt{2}} \left( \phi - \delta \right) .
\end{align}
If we insert this in eq.~\eqref{eq:Vphi1phi2}, we get
\begin{align}
 V(\phi, \delta) \, &= \, V_{\phi} +  V_{\delta},
\end{align}
with
\begin{align}
 V_\phi \, = \, V_0 \left\{ 1 + a \, \phi^2 + b \, \phi^4  \right\} \label{eq:VphiSimple}
\end{align}
and
\begin{align}
\frac{ V_{\delta}(\delta) }{ V_0 }  \, &= \,     
-\left(  2 \da \phi + 4 \db \phi^3  \right) \delta
+ \left(  a + 6b \phi^2 - 2b_{12} \phi^2 + 4d \phi^2  \right) \delta^2
- \underbrace{ 4 \db \phi \, \delta^3 }_{ \ll \, 4\db \phi^3 \delta }  + \underbrace{ b \, \delta^4 }_{\ll \, 6b \phi^2 \delta^2}
\notag \\
&\simeq -2\phi \left(  \da + 2 \db \, \phi^2  \right) \delta
+ \left[  a + \left(6b - 2b_{12}  + 4d  \right)\phi^2 \right] \delta^2.
\label{eq:Vdelta}
\end{align}
where $a$, $b$ describe the averaged potential for $\phi_1$ and $\phi_2$, while $\Delta a$, $\Delta b$ quantify the asymmetries between the potentials for $\phi_1$ and $\phi_2$:
\begin{align}
 a = \frac{a_1 + a_2}{2}, \quad b = \frac{b_1 + b_2 + b_{12}}{4}, \quad \da = \frac{a_2 - a_1}{2}, \quad \db = \frac{b_2 - b_1}{4}.
\end{align}
In these equations, $V_\phi$ contains the couplings along the perfectly D-flat direction, while $V_\delta$ contains the potential due to the inflaton's deviation from D-flatness.

The minimum of $\delta$ can be found by minimizing eq.~\eqref{eq:Vdelta}:
\begin{align}
\delta_{\text{min}} \, &\simeq \, \frac{ \da + 2\db \, \phi^2 }{ a + \left(  4d + 6b - 2b_{12}  \right)\phi^2 } \, \phi.
\end{align}
If we assume that $\delta$ tracks its minimum perfectly, then the effective inflaton potential is given by
\begin{align}
 V_{\text{eff}}(\phi) \, &= \, V(\phi, \delta_{\text{min}}) \, = \, V(\phi) + V_\delta(\delta_{\text{min}}).
\end{align}
The inflaton potential is changed by
\begin{align}
 V_\delta(\delta_{\text{min}})  ~ 
 &= ~  \, \frac{ -V_0 \left( \da + 2\db \, \phi^2 \right)^2 \phi^2 }{ a + \left(  4d + 6b - 2b_{12}  \right)\phi^2 } \, + \, O(\delta^3) ~
  \simeq ~  \frac{ -V_0 \left( \da + 2\db \, \phi^2 \right)^2 }{ 4d }.
\end{align}
With $d \sim O(10^{10})$ and $\da$, $\db \lesssim O(1)$, this correction is negligible compared to $V_\phi$, so assuming D-flatness is a very good approximation.

\emptyline

We should still check that $m^2_{\delta} > \Hubble^2 = V_0/3$ to justify our assumption that $\delta$ tracks its minimum:
\begin{align}
 \frac{ m^2_\delta }{ \Hubble^2 } \, &= \, \frac{3}{V_0} \, \frac{\partial^2 V}{\partial \delta^2} \, 
 = \, (\underbrace{ 6a + 36b \, \phi^2 - 12b_{12} \, \phi^2 }_{ \ll \, 1 } + 24d \, \phi^2)
\underbrace{ - 72 \db \, \phi \, \delta +  36 b \, \delta^2 }_{\ll \, 1} \, \simeq \, 24d \, \phi^2. \label{eq:deltamass}
\end{align}
For K\"{a}hler-driven tribrid inflation, the inflaton field value during inflation is bounded by $\phi^2 \gtrsim 10^{-6}$, which we can insert in eq.~\eqref{eq:deltamass} to find
\begin{align}
 \frac{ m^2_\delta }{ \Hubble^2 } \, \gtrsim \, 10^{5}.
\end{align}
Therefore, $m_\delta \gg \Hubble$, and our assumption that $\delta = \delta_{\text{min}}$ is warranted.

\emptyline

We conclude that any deviation from D-flatness during K\"{a}hler-driven tribrid inflation is completely negligible, and the inflaton potential indeed reduces completely to the simple single-field form \eqref{eq:VphiSimple} which has been previously studied in \cite{kahlerdriven}.

\section{Symmetries for the $A_4$ flavour model}
\label{appendix:symmetries}

The superpotential \eqref{eq:eqW} can be fixed by a set of suitable shaping symmetries in addition to the $A_4$ family symmetry. In particular, one can use an $U(1)_R$ symmetry, under which the superpotential has charge 2, and one $\mathbf{Z}_n$ symmetry for each flavon field.

\begin{table}[htb]
\centering
\begin{tabular}{ | l || c | c || c | c | c | c | c | c | c | }
  \hline
  & $A_4$ & $U(1)_{R}$ & $\textbf{Z}_{n_1}$ & $\textbf{Z}_{3}$ & $\textbf{Z}_{n_3}$ & $\textbf{Z}_{n_4}$ & $\textbf{Z}_{6}$ & $\textbf{Z}_{n_6}$ & $\textbf{Z}_{2n_7}$\\
  \hline
  $S_i$ & \textbf{1} & 2 &    \nocharge & \nocharge & \nocharge & \nocharge & \nocharge & \nocharge & \nocharge  \\
  $A_3$ & \textbf{3} & 2 &    \nocharge & \nocharge & $n_3-2$ & \nocharge & \nocharge & \nocharge & \nocharge  \\
  $A_4$ & \textbf{3} & 2 &    \nocharge & \nocharge & \nocharge & $n_4 - 2$ & \nocharge & \nocharge & \nocharge  \\
  $P_{34}$ & \textbf{1} & 2 &    \nocharge & \nocharge & $n_3 - 1$ & $n_4 - 1$ & \nocharge & \nocharge & \nocharge  \\
  $P_{36}$ & \textbf{1} & 2 &    \nocharge & \nocharge & $n_3 - 1$ & \nocharge & \nocharge & $n_6 - 1$ & \nocharge  \\
  $P_{46}$ & \textbf{1} & 2 &    \nocharge & \nocharge & \nocharge & $n_4 - 1$ & \nocharge & $n_6 - 1$ & \nocharge  \\
  $P_{35}$ & \textbf{1} & 2 &    \nocharge & \nocharge & $n_3 - 1$ & \nocharge & 5 & \nocharge & \nocharge  \\
  $P_{16}$ & \textbf{1} & 2 &    $n_1 - 1$ & \nocharge & \nocharge & \nocharge & \nocharge & $n_6 - 1$ & \nocharge  \\
  $P_{12}$ & \textbf{1} & 2 &    $n_1 - 1$ & $2$ & \nocharge & \nocharge & \nocharge & \nocharge & \nocharge  \\
  $D_{5}$ & \textbf{1} & 2 &    \nocharge & \nocharge & \nocharge & \nocharge & 2 & \nocharge & \nocharge  \\
  $D_{2}'$ & \textbf{1'} & 2 &    \nocharge & 1 & \nocharge & \nocharge & \nocharge & \nocharge & \nocharge  \\
  $D_{2}''$ & \textbf{1''} & 2 &    \nocharge & 1 & \nocharge & \nocharge & \nocharge & \nocharge & \nocharge  \\
  \hline
  $\Theta_1$ & \textbf{3} & 0 &    1 & \nocharge & \nocharge & \nocharge & \nocharge & \nocharge & \nocharge  \\
  $\Theta_2$ & \textbf{3} & 0 &    \nocharge & $\hphantom{11}1\hphantom{11}$ & \nocharge & \nocharge & \nocharge & \nocharge & \nocharge  \\
  $\Theta_3$ & \textbf{3} & 0 &    \nocharge & \nocharge & 1 & \nocharge & \nocharge & \nocharge & \nocharge  \\
  $\Theta_4$ & \textbf{3} & 0 &    \nocharge & \nocharge & \nocharge & 1 & \nocharge & \nocharge & \nocharge  \\
  $\Theta_5$ & \textbf{3} & 0 &    \nocharge & \nocharge & \nocharge & \nocharge & $\hphantom{11}1\hphantom{11}$ & \nocharge & \nocharge  \\
  $\Theta_6$ & \textbf{3} & 0 &    \nocharge & \nocharge & \nocharge & \nocharge & \nocharge & 1 & \nocharge  \\
  $\Theta_S$ & \textbf{1} & 0 &    \nocharge & \nocharge & \nocharge & \nocharge & \nocharge & \nocharge & 2  \\
  \hline
  $L$ & \textbf{3} & 1 &    \nocharge & \nocharge & \nocharge & \nocharge & \nocharge & \nocharge & \nocharge  \\
  $H_u$ & \textbf{1} & 0 &    \nocharge & \nocharge & \nocharge & \nocharge & \nocharge & \nocharge & 1  \\
  $H_d$ & \textbf{1} & 0 &    \nocharge & \nocharge & \nocharge & \nocharge & \nocharge & \nocharge & \nocharge  \\
  $N_1$ & \textbf{1} & 1 &    $n_1 - 1$ & \nocharge & \nocharge & \nocharge & \nocharge & \nocharge & $2n_7 - 1$  \\
  $N_2$ & \textbf{1} & 1 &    \nocharge & 2 & \nocharge & \nocharge & \nocharge & \nocharge & $2n_7 - 1$  \\
  $E_1$ & \textbf{1} & 1 &    \nocharge & \nocharge & \nocharge & $n_4 - 1$ & \nocharge & \nocharge & \nocharge  \\
  $E_2$ & \textbf{1} & 1 &    \nocharge & \nocharge & \nocharge & \nocharge & 5 & \nocharge & \nocharge  \\
  $E_3$ & \textbf{1} & 1 &    \nocharge & \nocharge & $n_3 - 1$ & \nocharge & \nocharge & \nocharge & \nocharge  \\
  \hline
\end{tabular}
\caption{One possible set of symmetries and charge assignments for the superpotential in eq.~\eqref{eq:eqW}.}
\label{tab:symmetryCharges}
\end{table}

The $U(1)_R$ charge assignment is straightforward. The mass parameters $\mu_i$ in the superpotential cannot be charged, so the $S_i$ must have 2 units of $U(1)_R$ charge. Then the $\Theta_i$ cannot have any $U(1)_R$ charge. Knowing this, we also find that the other auxiliary fields $A_i$, $P_{ij}$ and $D_i$ need 2 units of $U(1)_R$ charge. The only freedom we have is in the lepton sector, where we can distribute the $U(1)_R$ charge between the Higgs doublets $H_u$, $H_d$ and the lepton fields $L$, $N_i$, $E_i$. We choose to keep the Higgs fields uncharged and give 1 unit of $U(1)_R$ charge to each lepton.

To construct the $\textbf{Z}_{n}$ symmetries, we start with the observation that we want to keep the different flavons separate in the $S_i(\Theta_i^{n_i} - \mu_i^2)$ terms. For this reason, we start with one $\textbf{Z}_{k_i  n_i}$ symmetry for each flavon, with any set of positive integers $k_i$, under which the $i$-th flavon has charge $k_i$. To simplify the notation, we instead work with $k_i = 1$ and allow fractional charges for the other fields, except for the $A_4$ singlet flavon $\Theta_S$, which we normalize to 2 units of charge.

With the flavon charges fixed, we can uniquely determine the $\textbf{Z}_{n}$ charge assignments for the auxiliary fields $S_i$, $A_i$, $P_{ij}$ and $D_i$ by demanding that all terms in the superpotential must be allowed by the $\textbf{Z}_{n}$ symmetries.

The electron Yukawa couplings $\Theta L H_d E$ can always be allowed by choosing suitable charges for the $E_i$, which do not appear anywhere else. This fixes the $E_i$ charges.

Now the only remaining task is to make sure that the neutrino Yukawa coupling and mass terms are allowed by the $\textbf{Z}_{n}$ symmetries. There are several possible solutions, of which we have chosen a particularly simple one for table~\ref{tab:symmetryCharges}.

\end{document}